\documentclass[a4paper,fleqn,usenatbib]{mnras}
\usepackage[T1]{fontenc}
\usepackage{ae,aecompl}
\usepackage{graphicx}	% Including figure files
\usepackage{amsmath}	% Advanced maths commands
\usepackage{amssymb}	% Extra maths symbols
\usepackage{booktabs}	% Pretty tables for all!
\usepackage[normalem]{ulem}
\usepackage{url}

\usepackage{newtxtext,newtxmath}

\newcommand{\mga}{\textcolor{black} } % originally used to indicate internal changes

\usepackage{tablefootnote,footnotehyper} % footnotes in tables
\usepackage{array}
\newcolumntype{H}{>{\setbox0=\hbox\bgroup}c<{\egroup}@{}} % for hiding columns in tables

\title[KIC\,9773821]{A binary with a $\delta$\,Scuti star and an oscillating red giant: orbit and asteroseismology of KIC\,9773821}

\author[Simon J. Murphy et al.]{
Simon J. Murphy,$^{1,2}$\thanks{E-mail: simon.murphy@sydney.edu.au (SJM)}
Tanda Li,$^{3,2}$\thanks{E-mail: t.li.2@bham.ac.uk (TL)}
Sanjay Sekaran,$^{4}$
Timothy R. Bedding$,^{1,2}$
Jie Yu,$^{5}$\and
Andrew Tkachenko,$^{4}$
Isabel Colman,$^{1,2}$
Daniel Huber,$^{6}$
Daniel Hey,$^{1,2}$ \and
Tinatin Baratashvili,$^{4}$ and
Soetkin Janssens$^{4}$
\\
$^{1}$ Sydney Institute for Astronomy (SIfA), School of Physics, University of Sydney, NSW 2006, Australia\\
$^{2}$ Stellar Astrophysics Centre, Department of Physics and Astronomy, Aarhus University, 8000 Aarhus C, Denmark\\
$^{3}$ School of Physics and Astronomy, University of Birmingham, Edgbaston, B15 2TT, UK\\
$^4$ Instituut voor Sterrenkunde (IvS), KU Leuven, Celestijnenlaan 200D, B-3001 Leuven, Belgium\\
$^{5}$ Max-Planck-Institut f\"ur Sonnensystemforschung,  Justus-von-Liebig-Weg 3, 37077 G\"ottingen, Germany.\\
$^{6}$ Institute for Astronomy, University of Hawai`i, 2680 Woodlawn Drive, Honolulu, HI 96822, USA\\
}

\date{Accepted XXX. Received YYY; in original form ZZZ}
\pubyear{2015}
\begin{document}
\label{firstpage}
\pagerange{\pageref{firstpage}--\pageref{lastpage}}
\maketitle

\begin{abstract}
We study the $\delta$\,Scuti -- red giant binary KIC\,9773821, the first double-pulsator binary of its kind. It was observed by \textit{Kepler} during its four-year mission. Our aims are to ascertain whether the system is bound, rather than a chance alignment, and to identify the evolutionary state of the red giant via asteroseismology. An extension of these aims is to determine a dynamical mass and an age prior for a $\delta$\,Sct star, which may permit mode identification via further asteroseismic modelling. We determine spectroscopic parameters and radial velocities (RVs) for the red giant component using HERMES@Mercator spectroscopy. Light arrival-time delays from the $\delta$\,Sct pulsations are used with the red-giant RVs to determine that the system is bound and to infer its orbital parameters, including the binary mass ratio. We use asteroseismology to model the individual frequencies of the red giant to give a mass of \mga{$2.10^{+0.20}_{-0.10}$\,M$_{\sun}$ and an age of $1.08^{+0.06}_{-0.24}$\,Gyr}. We find that it is a helium-burning secondary clump star, confirm that it follows the standard $\nu_{\rm max}$ scaling relation, and confirm its observed period spacings match their theoretical counterparts in the modelling code {\sc mesa}.  Our results also constrain the mass and age of the $\delta$\,Sct star. We leverage these constraints to construct $\delta$\,Sct models in a reduced parameter space and identify four of its five pulsation modes.
\end{abstract}


%%%%%%%%%%%%%%%%%%%%%%%%%%%%%%%%%%%%%%%%%%%%%

%%%%%%%%%%%%%%%%% BODY OF PAPER %%%%%%%%%%%%%%%%%%

\section{Introduction}

Double-lined spectroscopic binaries (SB2s) provide critical tests for stellar evolution \citep{polsetal1997,pourbaix2000,claret2007,deminketal2007,lastennet&valls-gabaud2002}. Their value lies in the multitude of information that can be extracted from the composite spectra, such as the individual stellar effective temperatures, the system metallicity, and the radial velocities (RVs) whose variations encode the dynamical masses and mass-ratio of the stars \citep{vogel1889,pickering1890,stebbins1911}. These RVs can further be used to infer the binary orbital parameters, whose distribution functions feed directly into our understanding of binary star formation \citep{duchene&kraus2013,moe&distefano2017,murphyetal2018,shahaf&mazeh2019}, such as the spatial scales at which fragmentation of protostellar disks outcompetes fragmentation of molecular cores as the dominant binary star formation mechanism \citep{tohline2002,bate2009,kratter2011,tobinetal2016}.

In parallel with spectroscopy, asteroseismology has established itself as an indispensable tool for astrophysical investigation \citep{aertsetal2010}. For example, catalogues of stellar properties that combine spectroscopy and asteroseismology for red giants \citep{pinsonneaultetal2014,pinsonneaultetal2018} have greatly informed Galactic dynamics \citep[e.g.][]{spitonietal2019,hasselquistetal2020,haydenetal2020,ratcliffeetal2020}, thanks largely to the high age precision typically achievable \citep{chaplin&miglio2013,lebretonetal2014,hekker&christensen-dalsgaard2017,basu&hekker2020}.

Masses and radii for red giants are often determined from so-called asteroseismic scaling relations \citep{stelloetal2008,kallingeretal2010}, using the observed frequency of maximum power, $\nu_{\rm max}$, and large frequency spacing, $\Delta\nu$. Since $\Delta\nu$ scales with the square-root of the mean density of the star \citep{ulrich1986}, the evolutionary stage of red giants -- that is, the extent of their ascension up the red giant branch (RGB) -- is readily discernible from $\Delta\nu$ alone. However, there is another phase of red giant evolution that occurs after the helium flash known as the red clump (RC), during which stars have a similar density and their evolutionary tracks overlap with the RGB. Except for metal-poor stars, the core-helium burning RC stars are observationally indistinguishable from the hydrogen-shell burning RGB stars by surface properties alone, but in each phase the gravity modes have different period spacings due to the different properties of the core \citep{beddingetal2011,mosseretal2012a,vrardetal2016}, allowing the two to be distinguished asteroseismically. Stars above $\sim$1.8\,M$_{\odot}$, however, do not undergo a helium flash and these stars sit in the so-called secondary clump, whose period spacings are similar to RGB counterparts of slightly higher mass \citep[e.g.][]{stelloetal2013a,mosseretal2014,bossinietal2017}. This further challenges the identification of their evolutionary status, and poses some important questions, such as whether scaling relations apply to secondary clump stars \citep{miglioetal2012,sharmaetal2016}. To analyse potential secondary clump stars, external constraints are typically required, such as priors on age or mass. Those priors are often obtainable in binary systems.

By comparison, asteroseismology of $\delta$\,Sct stars is less advanced, with a major barrier being mode identification \citep{breger2000,bowman&kurtz2018}. Pairs of low-radial-order (low $n$) radial modes can have period ratios that enable those modes to be identified \citep{petersen&christensen-dalsgaard1996}, but not every mode in $\delta$\,Sct stars is driven to an observable amplitude, so those modes might not be present. A recent breakthrough has been the discovery that some $\delta$\,Sct stars pulsate in regular patterns that form ridges in an \'echelle diagram \citep{beddingetal2020}. This allows the modes to be identified somewhat analogously to solar-like oscillations, but is only possible for some stars close to the zero-age main-sequence, including on the pre-main-sequence \citep{murphyetal2020e}. Binary systems can also be of assistance in analysing $\delta$\,Sct stars via the provision of dynamical masses, which may lead to sufficiently constrained models to allow the observed modes to be identified \citep[e.g.][]{streameretal2018,steindletal2020}.

\begin{figure*}
\begin{center}
\includegraphics[width=0.98\textwidth]{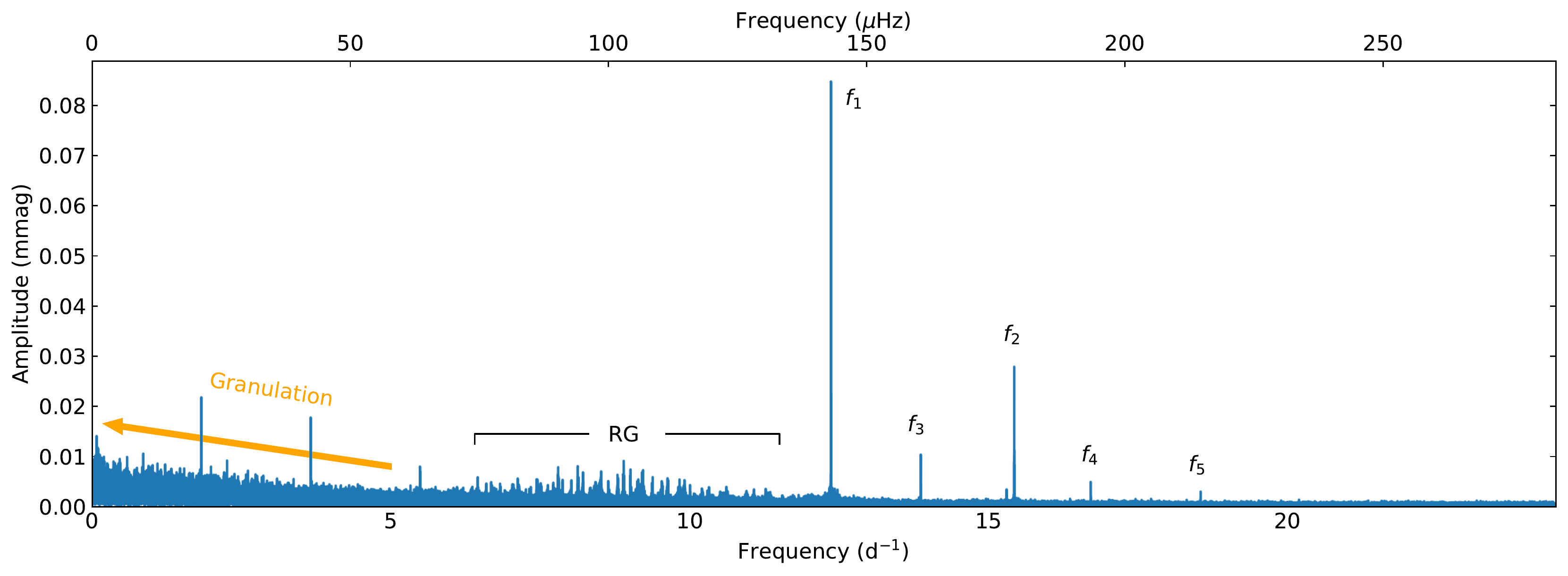}
\caption{The amplitude spectrum of the long-cadence \textit{Kepler} light-curve of KIC\,9773821, out to the Nyquist frequency of 24.49\,d$^{-1}$ (283.45\,$\upmu$Hz). The labelled frequencies $f_1$ -- $f_5$ originate in the $\delta$\,Sct component, while the red giant contributes a broad power excess near 8.6\,d$^{-1}$ (100\,$\upmu$Hz) and a granulation background. The top axis shows frequencies in $\upmu$Hz.}
\label{fig:FT}
\end{center}
\end{figure*}

In this paper we analyse KIC\,9773821, whose light curve shows the pulsations of both a $\delta$\,Sct and a red giant star (Fig.\,\ref{fig:FT} shows its amplitude spectrum). Such double-pulsator binaries are named PB2s in analogy to the SB2s of spectroscopy \citep{murphyetal2016b}. 
%Our aims are to ascertain whether the system is bound, rather than a chance alignment, and to identify the evolutionary state of the red giant via asteroseismology. An extension of these aims is to determine a dynamical mass and an age prior for a $\delta$\,Sct star, which may permit mode identification via further asteroseismic modelling.
% The $\delta$\,Sct pulsations, which dominate the light curve, have detectable shifts in their observed phases that we use to infer the orbital motion. We also obtained spectra, which are dominated by the red giant, from which we measure RV variations. In this way, we confirm the two components are bound, we calculate their mass ratio, and we use the binary orbital parameters to guide asteroseismology of the system.
Throughout our investigations we adopt the convention that the $\delta$\,Sct star is component 1 and the RG is component 2 (e.g. in the subscripts for RV amplitudes, $K_1$ and $K_2$). This is because our investigation began with the \textit{Kepler} light curve, whose variability is dominated by the $\delta$\,Sct star (Fig.\,\ref{fig:FT}), and because light travel time variations were discovered in the $\delta$\,Sct pulsations, providing the first evidence that the stars may be bound. However, the red giant is the more luminous, more evolved and more massive object.

The {\it Kepler} observations are described in Sec.\,\ref{sec:kep}, demonstrating the double-pulsator nature of the system. Our spectroscopic observations and analysis, including radial velocity extraction, are described in Sec.\,\ref{sec:spectro}. In Sec.\,\ref{sec:orbit} we measure light travel time variations for the $\delta$\,Sct star, and combine them with the radial velocities to show that the stars have the same orbital period, and are thus a bound system. The orbital analysis of Sec.\,\ref{sec:orbit} guides deeper asteroseismology of the red giant (Sec.\,\ref{sec:seismo}) to infer its evolutionary state, and the $\delta$\,Sct star (Sec.\,\ref{sec:dsct_models}) to identify its modes. Our conclusions are given in Sec.\,\ref{sec:conclusions} and our data are given in an appendix.

\section{Kepler observations}
\label{sec:kep}

KIC\,9773821 (magnitude $K_p = 9.86$) has 4\,yr of \textit{Kepler} observations in long-cadence mode (29.45-min sampling), for a total duration of $\Delta T$ = 1470.47\,d. The corresponding frequency resolution is $1/(\Delta T) = 0.00068$\,d$^{-1}$ (0.0079\,$\upmu$Hz). We used the light curve processed with the `msMAP' pipeline \citep{stumpeetal2014}, which we downloaded from KASOC.\footnote{\url{https://kasoc.phys.au.dk/}} We converted the fluxes to magnitudes and subtracted the mean magnitude from each of the 18 \textit{Kepler} `quarters', including the commissioning quarter, Q0.

The Fourier transform of the \textit{Kepler} lightcurve is shown in Fig.\,\ref{fig:FT}, which shows both $\delta$\,Sct and red-giant oscillations. Five $\delta$\,Sct peaks are labelled, with their extracted frequencies and amplitudes given in Table\:\ref{tab:freqs}; we describe these in Sec.\,\ref{ssec:dsct}. The broad power excess near 8.6\,d$^{-1}$ (100\,$\upmu$Hz) consists of red-giant (RG) oscillations and is analysed in Sec.\,\ref{ssec:rg}. Other key features in the Fourier transform are a rising background at low frequency, which is a typical feature of red-giant power spectra that arises from granulation, and three low frequencies starting at 1.8\,d$^{-1}$ (21\,$\upmu$Hz) that are almost harmonics of each other (Sec.\,\ref{ssec:lowf}). Although only long-cadence observations are available, barycentric corrections to the time-stamps of the data result in a modulated sampling rate as the \textit{Kepler} spacecraft orbits the Sun, allowing us to assert that none of the extracted peaks are Nyquist aliases \citep{sna}.

The oscillations of both stars are affected by the orbital motion, which causes periodic Doppler shifts of each star's pulsations. The effect in Fourier space is the generation of a frequency multiplet \citep{shibahashi&kurtz2012} which can be used to extract orbital information if the oscillations are coherent \citep{shibahashietal2015}. However, orbital information cannot be extracted from the low-amplitude, stochastic red giant oscillations \citep{comptonetal2016}. 
%To sharpen up the red-giant oscillation spectrum prior to its analysis, and for this purpose only, we corrected the time-stamps of the data to the rest-frame of the red giant, using the orbital analysis in Sec.\,\ref{sec:orbit}. The effect is shown in Fig.\,\ref{fig:rg_sharpened}.

\begin{table}
\centering
\caption{Pulsation frequencies and amplitudes from the $\delta$\,Sct component.}
\label{tab:freqs}
\begin{tabular}{c c c r}
\toprule
ID & Frequency & Frequency & Amplitude\\
 & d$^{-1}$ & $\upmu$Hz & $\pm$1\,$\upmu$mag \\
\midrule
$f_1$ & 12.35999 & 143.0554 & 85 \\
$f_2$ & 15.42488 & 178.5287 & 28 \\
$f_3$ & 13.86231 & 160.4434 & 10 \\
$f_4$ & 16.70400 & 193.3333& 5 \\
$f_5$ & 18.54437 & 214.6339 & 3 \\
\bottomrule
\end{tabular}
\end{table}

%\begin{figure}
%\begin{center}
%\includegraphics[width=0.48\textwidth]{figures/rg-times.pdf}
%\caption{The {\it Kepler} long-cadence amplitude spectrum covering the frequency range of the red giant oscillations, with and without corrections of the observation times to the rest frame of the red giant.}
%\label{fig:rg_sharpened}
%\end{center}
%\end{figure}

\subsection{The $\delta$\,Sct pulsations}
\label{ssec:dsct}

The $\delta$\,Sct pulsations appear to have some regularity, which led us to search for signs of harmonics or sum frequencies (e.g. \citealt{papics2012,kurtzetal2015a}), but we found none. However, there are some interesting patterns among the five $\delta$\,Sct peaks. Firstly, $f_3$ lies almost exactly halfway between $f_1$ and $f_2$. Equidistance would correspond to a frequency of 13.892\,d$^{-1}$, whereas $f_3$ lies 0.03\,d$^{-1}$ away at 13.862\,d$^{-1}$, which is 44 times the frequency resolution and suggests this is not a combination frequency. Similarly, $f_5$ lies 0.054\,d$^{-1}$ away from being the combination $2f_2 - f_1 = 18.490$\,d$^{-1}$. We conclude that these near-combinations are coincidences and we explore identities for these modes in our $\delta$\,Sct modelling (Sec.\,\ref{sec:dsct_models}).

\subsection{The red-giant oscillations}
\label{ssec:rg}

\begin{figure*}
\begin{center}
\includegraphics[width=0.48\textwidth]{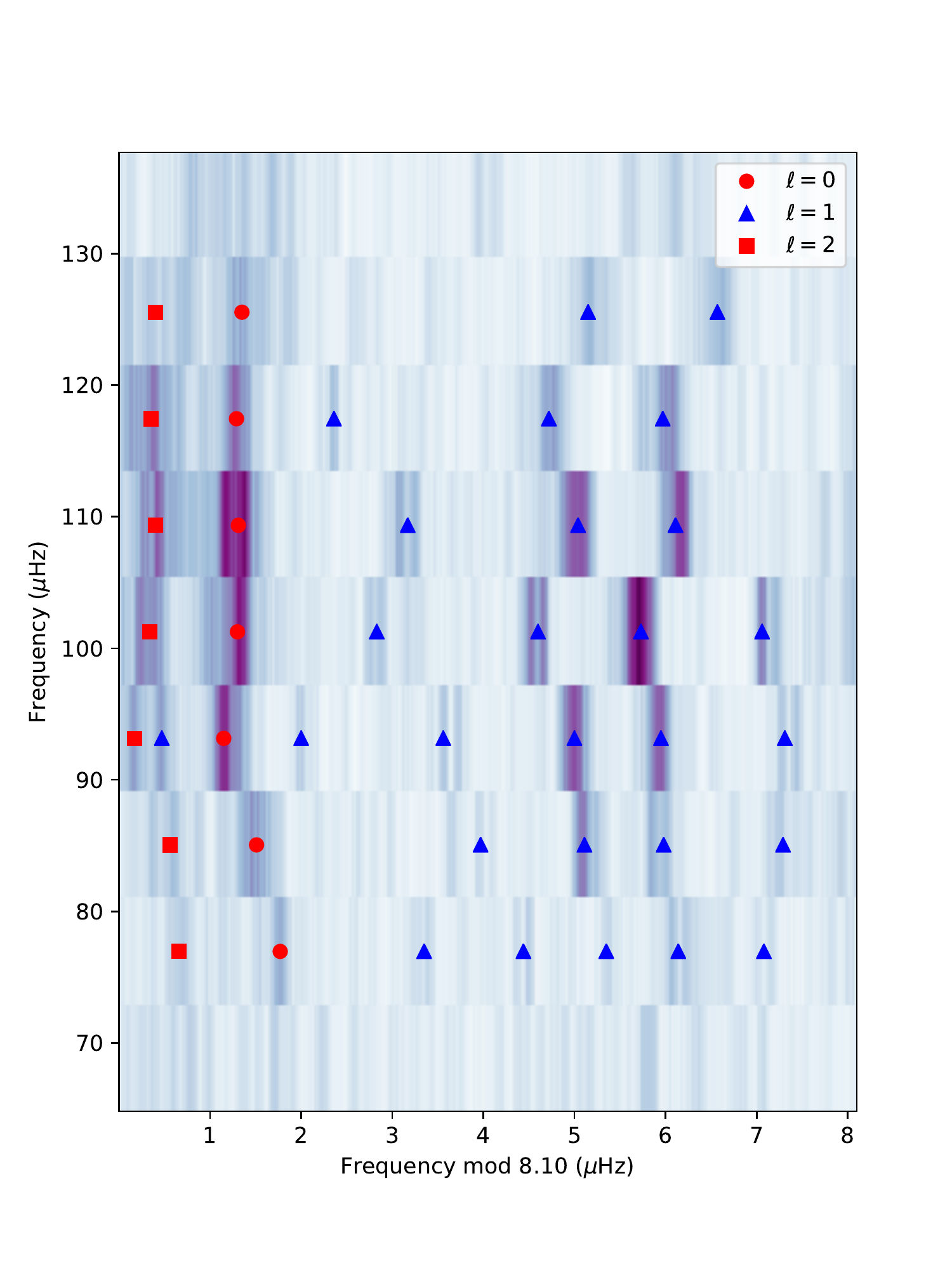}\includegraphics[width=0.48\textwidth]{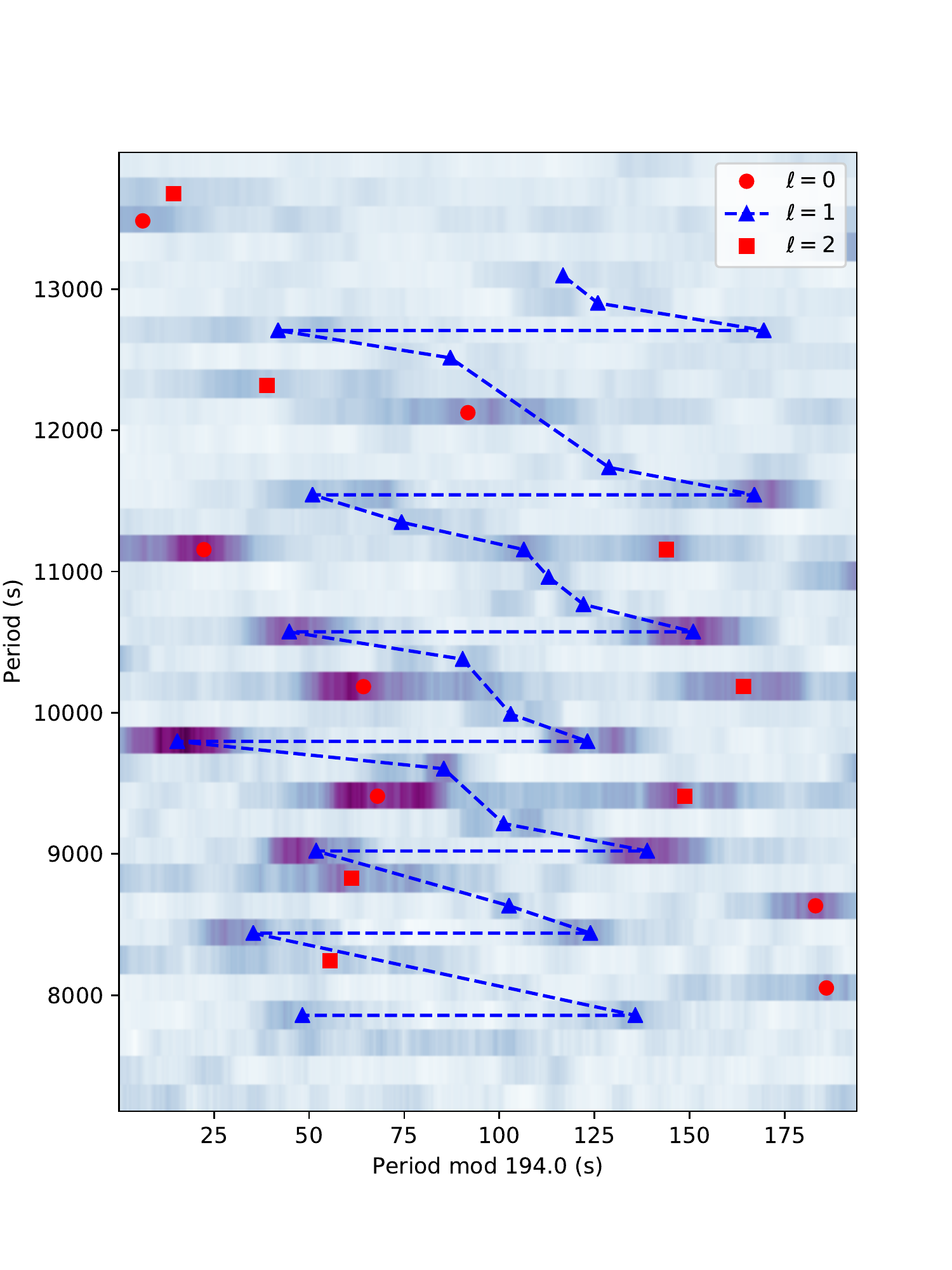}\\
\caption{\'Echelle diagrams in frequency (\textbf{left}) and period (\textbf{right}) covering the RG oscillations of KIC\,9773821. The greyscale shows the observed amplitude spectrum and the symbols identify the modes that were used in modelling.}
\label{fig:echelle}
\end{center}
\end{figure*}

The oscillations from the red giant are clearly visible in the power spectrum with very high signal-to-noise, making it straightforward to measure and identify dozens of modes.  The left panel of Fig.\,\ref{fig:echelle} shows the power spectrum in \'echelle format, with a large frequency separation of $\Delta\nu = 8.05\,\upmu$Hz, calculated from the autocorrelation of the power spectrum.  We can easily recognise the patterns of modes with degrees $\ell=0$ (red circles), $\ell=1$ (blue squares) and $\ell=2$ (red triangles).  The right panel shows the period \'echelle diagram \citep{beddingetal2011}, and we see the characteristic pattern of $\ell=1$ modes. By aligning the structure of these \mbox{$\ell=1$} modes vertically we determined an asymptotic period spacing \mga{$\Delta\Pi = 194.0$}\,s. We used this pattern to guide the selection of the $\ell=1$ modes.

The extracted frequencies, their uncertainties, and their and mode identifications are given in Table\:\ref{tab:RG_freqs}. We measured the frequencies after smoothing the power spectrum slightly in order to smooth modes broadened by the finite lifetimes.  Some of the $\ell=1$ peaks are split into multiplets by rotation.  Our asteroseismic analysis (Sec.\,\ref{ssec:oscillation_models}) is based on non-rotating models and, by measuring peaks in the smoothed spectrum, we measured the central ($m=0$) component. Uncertainties on the measured frequencies are generally lower for stronger peaks and we were guided by the equations in \citet{montgomery&odonoghue1999} and \citet{kjeldsen&bedding2012}.  In order to be conservative, we adopted an uncertainty of 0.03\,$\upmu$Hz for most modes.  In practice, the actual values of the uncertainties do not influence the results of the modelling and those given in Table\:\ref{tab:RG_freqs} are intended to be approximate values.

%When modelling, we also add a systematic uncertainty to the observational values (Sec.\,\ref{ssec:oscillation_models}).

We measured the frequency of maximum power, $\nu_{\rm max}$, in a two-step process using the SYD pipeline \citep{huberetal2009}. First, the background of a power-density spectrum was modelled by a sum of two power laws and a white-noise component, and then subtracted from the original power-density spectrum. This spectrum was subsequently smoothed using a Gaussian with a full width at half-maximum of 2($\Delta\nu$) and $\nu_{\rm max}$ was taken as the peak of the heavily smoothed spectrum. Its uncertainty was calculated by perturbing the original power-density spectrum 500 times with a $\chi^2$ distribution with two degrees of freedom and repeating the above procedure on each perturbed spectrum. The standard deviation of the resulting distribution was adopted as the formal uncertainty \citep{huberetal2011}. In this way, we calculated $\nu_{\rm max}=102.06\pm0.96\,\upmu$Hz.

%%% In the extraction and identification of the red-giant oscillations (Fig.\,\ref{fig:echelle}), we err on the side of caution where identification is ambiguous. In particular, observational mode identification was performed independently of the modelling because uncertainties in the physics of the models, including from surface corrections, all lead to uncertainties in the calculated oscillation modes. It is therefore best if model predictions are not used to aid the observational mode identification. In addition, we preferred a realistic uncertainty on the computed stellar components, so any ambiguous modes, such as the two predicted $\ell=1$ modes that lie between the $\ell=0$ and 2 modes around 100\,$\upmu$Hz, were left out of the observed identifications. 
%% actually, I think it is good to use models to help identify modes, we often do that (but not in this case, since we didn't have models when I did the mode IDs)

\subsection{Low-frequency peaks}
\label{ssec:lowf}

The low frequency peaks at 1.8292, 3.6578, and 5.4860\,d$^{-1}$ (21.171, 42.336, and 63.495\,$\upmu$Hz; Fig.\,\ref{fig:FT}) are not exact harmonics of each other and are broader than the frequency resolution of the data. It is therefore unlikely that they result from ellipsoidal variability and reflection effects in a compact binary \citep{colmanetal2017}.  \citet{bowman2017} discussed similar features in other stars, which he called `organ pipe stars', and argued against a rotational origin unless significant latitudinal differential rotation is present in A-type stars. Since these low frequencies are also dissimilar to the spacing of the $\delta$\,Sct p\:modes in KIC\,9773821, their origin remains unclear.

\section{Spectroscopy}
\label{sec:spectro}

\begin{figure}
\begin{center}
\includegraphics[width=\hsize]{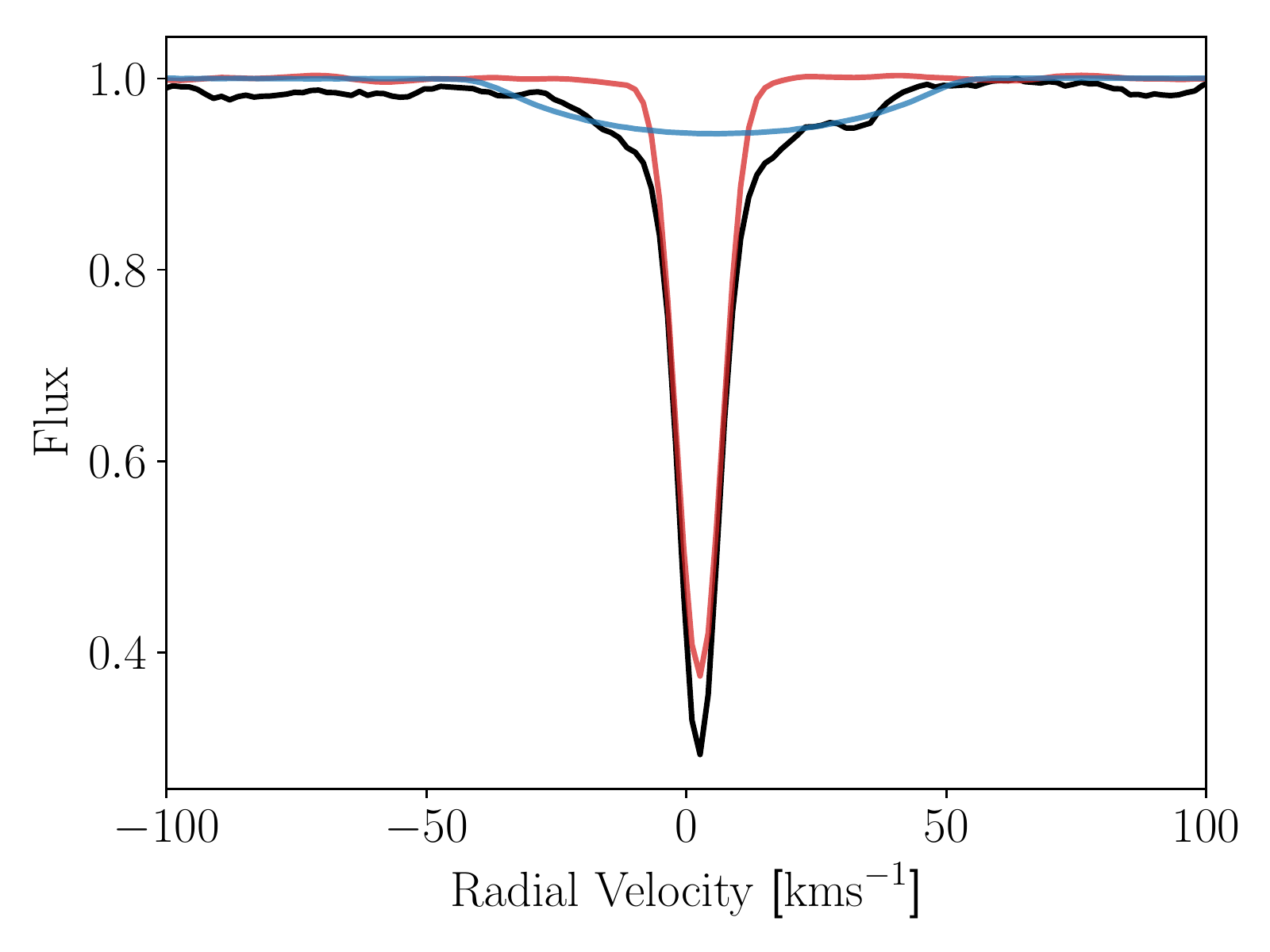}
\caption{One of the observed (black) LSD profiles of KIC97732832. The sharp-lined RG and the broad-lined $\delta$\,Sct profile are indicated by the red and blue synthetic LSD profiles calculated from synthetic spectra generated using \textsc{gssp}.}
\label{fig:LSD}
\end{center}
\end{figure}

We obtained high-resolution ($R\sim 85000$) spectroscopy of the system using the HERMES spectrograph \citep{Raskin2011} attached to the 1.2-m Mercator telescope at the Roque de los Muchachos observatory on La Palma, Spain. In total, 17 spectra of KIC\,9773821 at a $\text{S/N}\sim80$ were obtained from July 2012 to October 2019. We computed Least-Squares Deconvolution (LSD) profiles (\citealt{Donati1997}, as implemented by \citealt{Tkachenko2013b}) from the normalised spectra to determine if the radial velocities of the individual components could be extracted. We found that the profiles of the individual components were blended at all orbital phases, which is unsurprising for a long-period binary ($P_{\text{orb}}=482$\,d; Sec.\,\ref{sec:orbit}). In addition, the more-luminous slowly-rotating red giant dominates the LSD profiles, preventing us from extracting reliable RVs for the more rapidly-rotating $\delta$\,Sct component (Fig.\,\ref{fig:LSD}). RVs for the red giant were extracted to a precision of $\sim$0.15\,km\,s$^{-1}$ (Table\:\ref{tab:RG_RVs}).

We determined the atmospheric parameters by first separating the component spectra into their individual components using the spectral disentangling technique \citep{SS1994,Hadrava1995}. We used the Fourier domain-based disentangling code FDBinary \citep{Ilijic2004b}, which allowed us to simultaneously optimise the orbital parameters of the system while obtaining the component spectra. Allowing both $K_{1}$ and $K_{2}$ to vary in the disentangling resulted in unphysical component spectra, and we therefore had to fix $K_{1}$ at the value 23.91\,km\,s$^{-1}$ determined from phase modulation (described in Sec.\,\ref{sec:orbit}). After $K_{1}$ was fixed, we obtained realistic component spectra.

The atmospheric parameters were determined by fitting synthetic spectra to the 4800--5700~\AA~region\footnote{The normalised continuum level drops below unity for red giant stars at wavelengths lower than $\sim4800$~\AA\ due to the Balmer jump, requiring an iterative spectral normalisation approach and with increased errors that would inevitably propagate into the red giant component spectrum.} using the Grid Search in Stellar Parameters (\textsc{gssp}) software package \citep{tkachenko2015}. \textsc{gssp} is able to generate and fit synthetic spectra to an observed spectrum using the \textsc{SynthV} radiative transfer code \citep{tsymbal1996} combined with a grid of atmospheric models from the \textsc{LLmodels} code \citep{Shulyak2004}. Using this method, we determined the atmospheric parameters $T_{\text{eff}}$, log $g$, microturbulent velocity ($v_{\text{mic}}$), macroturbulent velocity ($v_{\text{mac}}$), projected rotational velocity ($v$ sin $i$), and the light contribution for each star, as well as the global metallicity ($\mathrm{[M/H]}$). We determined uncertainties from the distribution of $\chi^{2}$ values of the fit of each synthetic spectrum to the observed spectrum. Due to the degeneracy between $\mathrm{[M/H]}$ and the light ratio, we would have to place constraints on at least one of the stellar luminosities ($L_{\text{RG}}$ or $L_{\rm \delta\,Sct}$) to obtain $\mathrm{[M/H]}$ directly. Nonetheless, we note that the inferred metallicity of each component was equal for light ratios $L_{\text{RG}}$/$L_{\rm \delta\,Sct}=3.2\pm0.3$, and they rapidly diverge outside this range. To obtain individual luminosity constraints, the atmospheric parameters were determined iteratively with the asteroseismic analysis of the red giant (Sec.\,\ref{sec:seismo}), starting with the best-fitting $T_{\rm eff}$ of $5080\pm130$\,K determined at the (initially fixed) solar metallicity. The $\delta$\,Sct parameters were determined thereafter, and include a 2\% systematic uncertainty on the effective temperature \citep{casagrandeetal2014,whiteetal2018}.

\section{Orbital analysis}
\label{sec:orbit}

We applied the phase modulation method \citep{murphyetal2014,murphy&shibahashi2015} to measure light arrival-time delays for the $\delta$\,Sct pulsations. The method works by measuring the pulsation frequencies as precisely as possible using the entire 4-yr \textit{Kepler} data set, then subdividing the light curve and measuring the pulsation phases in each subdivision whilst keeping the frequencies fixed. Phase shifts are readily converted into time delays, which contain information on the star's position in its orbit. When the $\delta$\,Sct star is on the far side of the orbit, the arrival of its light at Earth is delayed, whereas light arrives early (a negative time delay) when the star is at the near side of its orbit. Each pulsation frequency should respond in the same way, that is, should give the same time delay curve, if the modulation is due to binary motion rather than intrinsic phase modulation. The method has been applied to hundreds of $\delta$\,Sct pulsators, whose coherent pulsations appear to be good `clocks' in this regard \citep{murphyetal2018}.

A free parameter in phase-modulation analyses is the subdivision size. Longer subdivisions result in more precise phases and thus smaller uncertainties per point but will undersample the orbital signal if the orbital period is short. \citet{heyetal2020a} developed a forward-modelling method to fit orbits to the light curve directly, which offers great improvement for short-period binaries \mbox{($P\lesssim100$\,d).} For longer periods as in this case, the subdividing approach remains sufficiently accurate and we used it for this analysis. Following application of the subdividing approach, we confirmed our results with the {\sc maelstrom} forward-modelling code \citep{heyetal2020b}.

\begin{figure*}
\begin{center}
\includegraphics[width=0.98\textwidth]{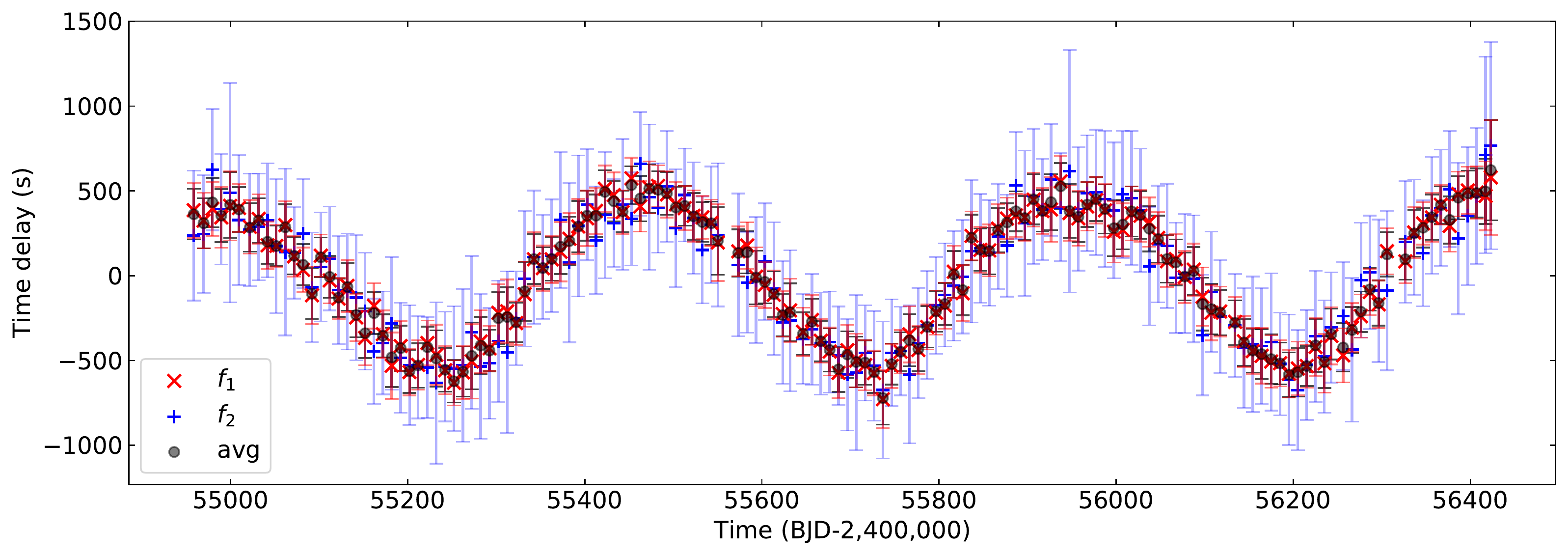}
\caption{Time delays for the strongest two $\delta$\,Sct frequencies, and their weighted average, using 10-day subdivisions. The periodic variation contains the orbital information.}
\label{fig:tds}
\end{center}
\end{figure*}

We extracted time delays for $f_1$ and $f_2$, which are the strongest of the five extracted peaks (Table\:\ref{tab:freqs}). The next-strongest peak, $f_3$, is not statistically significant in 10-d subdivisions of the light curve, and $f_4$ and $f_5$ are weaker still. However, $f_3$ is significant in 20-d subdivisions, and shows time delays consistent with $f_1$ and $f_2$, albeit with large scatter. Fig.\,\ref{fig:tds} shows the time delays with their weighted average (weighting by the inverse square of the pulsation phase uncertainties), indicating a projected light travel time variation of $\pm$500\,s, corresponding to a projected semi-major axis ($a_1 \sin i$) of approximately 1\,au. Note that the projected separation of the two stars, $a \sin i$, is larger than this [$a \sin i = (a_1 + a_2)\sin i$], and note also that time-delay uncertainties are overestimated because there is considerable variance in the data from the red-giant oscillations.

\begin{figure*}
\begin{center}
\includegraphics[width=0.98\textwidth]{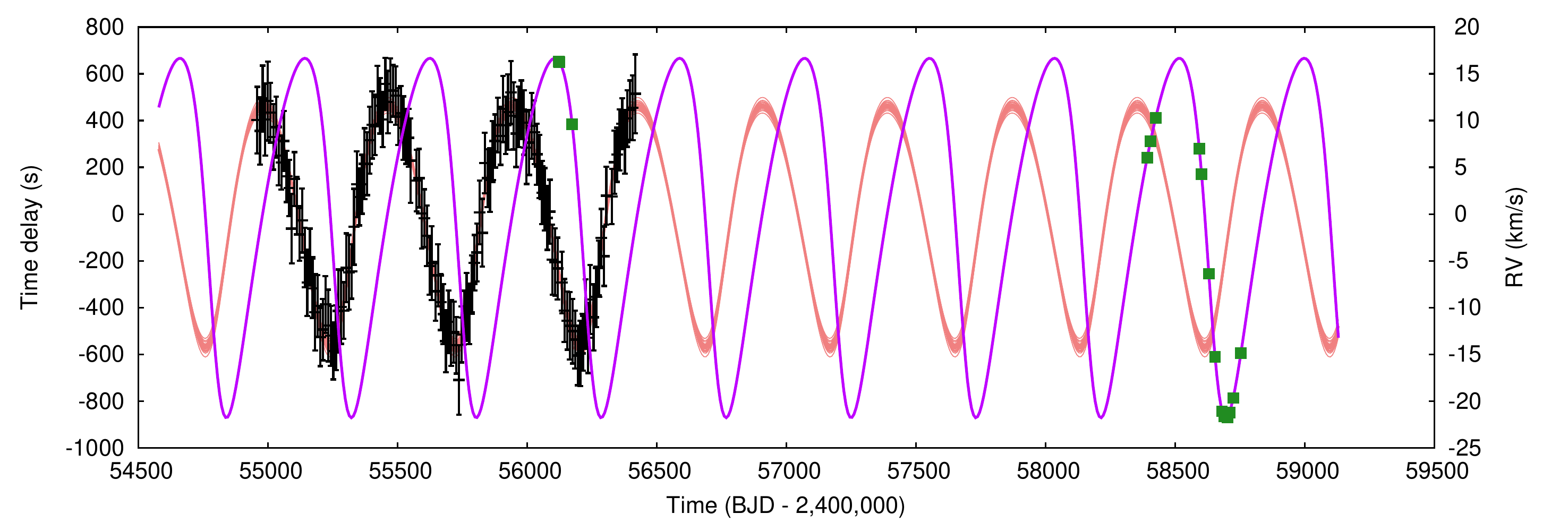}
\caption{Orbit of KIC\,9773821. The black data with error bars are the weighted-average time delay measurements from \textit{Kepler} observations of the two $\delta$\,Sct modes (left y-axis). The red lines are 25 random samples of the Markov chain of its orbit. The green squares, whose error bars are smaller than the plot symbols, are the HERMES@Mercator RV data for the red-giant secondary (right y-axis). The magenta lines are the corresponding 25 random samples for the red-giant's (RV) orbit, computed simultaneously with the $\delta$\,Sct star's time-delay orbit, using the same system parameters. The resulting parameters are given in Table\:\ref{tab:orbit}.}
\label{fig:orbit}
\end{center}
\end{figure*}

Radial velocities, RV(t), are the time derivative of the time delays, $\tau(t)$,
\begin{eqnarray}
RV(t) = {\rm c} \frac{d\tau(t)}{dt},
\end{eqnarray}
and depend on the same orbital parameters. \citet{murphyetal2016b} verified the PM method by comparison with RVs, and those authors and others \citep{lampensetal2018,derekasetal2019} have presented joint modelling of RVs and time delays, similar to the one we perform here for KIC\,9773821. A key difference here is that the RV curve belongs to the red giant and so is in exact anti-phase with that calculated from the time delays of the $\delta$\,Sct star (Fig.\,\ref{fig:orbit}). This proves that the two stars orbit each other and are not a chance alignment.

We determined the orbital parameters using the Markov chain Monte Carlo method based on the Metropolis-Hastings algorithm, as described by \citet{murphyetal2016b}. In each iteration of the Markov chain the two components, as described by their respective time-delay and RV curves, had the same orbital period and eccentricity, while their longitudes of periastron $\varpi$ differed by precisely $\uppi$ radians. Their projected semi-major axes were sampled via independent proposals at each iteration in the Markov chain, and the inverse ratio of those axes gives the mass ratio $q$ of the system, that is,
\begin{eqnarray}
q = \frac{M_2}{M_1} = \frac{a_1 \sin i}{a_2 \sin i }.
\end{eqnarray}
The total $\chi^2$ was used to evaluate each iteration, and no additional weights were applied to the time-delay or RV data sets. 
%We also calculated the reduced chi-squared, $\chi^2/N$, where $N$ is the number of data points, for the time-delay and the RV datasets. Its value was approximately three times larger for the time delays than than for the RVs, but there are fewer RV data overall so reweighting was not deemed necessary. 
The chain was run for 25\,000 iterations and manually checked for convergence, after which the total $\chi^2$ was 39.8 from 143 time delays and 17 RV data. All orbital parameters were determined as the medians of their marginalised posteriors, and the uncertainties were the 15.9 and 84.1 percentiles of those posteriors, together bracketing the central 68.2\:per\:cent of the data. Randomly-drawn samples from the converged chain are shown in Fig.\,\ref{fig:orbit} and the orbital parameters are given in Table\:\ref{tab:orbit}.

Since the mass ratio, the mass function, and one of the individual component masses are known precisely (the modelled red giant mass for the secondary clump scenario, discussed in Sec.\,\ref{sec:seismo}), it is possible to calculate the orbital inclination. To do this, we performed a 100\,000-iteration Monte Carlo simulation involving each of the above parameters to sample $\sin i$ according to
\begin{eqnarray}
f(m_1, m_2, \sin i) = f_M = \frac{m_2^3 \sin^3 i}{(m_1 + m_2)^2}
\end{eqnarray}
which can be rewritten as
\begin{eqnarray}
\sin^3 i = \frac{f_M}{m_2} (1 + 1/q)^2.
\end{eqnarray}
The resulting stellar inclination, $i$, was \mga{$81^{+9}_{-10}$}\,deg, meaning the system is observed nearly edge-on.

\begin{table}
\centering
\caption{Orbital parameters for the KIC\,9773821 system. $K_1$ and $K_2$ are calculated from the other orbital parameters, where star 1 is the $\delta$\,Sct star and star 2 is the red giant.}
\label{tab:orbit}
\begin{tabular}{c c r@{}l}
\toprule
\multicolumn{1}{c}{Parameter} & Units & \multicolumn{2}{c}{Values}\\
\midrule
\vspace{1.5mm}
$P_{\rm orb}$ & d & $481.93$&$^{+0.13}_{-0.12}$\\
\vspace{1.5mm}
$e$ & -- & $0.241$&$^{+0.0030}_{-0.0031}$\\
\vspace{1.5mm}
$\varpi$ & rad & $2.154$&$^{+0.014}_{-0.011}$\\
\vspace{1.5mm}
$t_{\rm p}$ & d & $2\,455\,273.9$&$^{+1.5}_{-1.3}$\\
\vspace{1.5mm}
$\gamma$ & km\,s$^{-1}$ & \multicolumn{2}{c}{8.86}\\
\vspace{1.5mm}
$f(m_1,m_2,\sin i)$ & M$_{\odot}$ & $0.623$&$^{+0.061}_{-0.053}$\\
\vspace{1.5mm}
$f(m_2,m_1,\sin i)$ & M$_{\odot}$ & $0.324$&$^{+0.031}_{-0.028}$\\
\vspace{1.5mm}
$a_1 \sin i / c$ & s & $513$&$^{+16}_{-15}$\\
\vspace{1.5mm}
$a_2 \sin i / c$ & s & $412.1$&$^{+1.0}_{-1.1}$\\
\vspace{1.5mm}
$a_1 \sin i$ & au & $1.03$&$^{+0.03}_{-0.03}$\\
\vspace{1.5mm}
$a_2 \sin i$ & au & $0.8259$&$^{+0.0020}_{-0.0022}$\\
\vspace{1.5mm}
$K_1$ & km\,s$^{-1}$ & $23.91$&$^{+0.75}_{-0.70}$\\
\vspace{1.5mm}
$K_2$ & km\,s$^{-1}$ & $19.21$&$^{+0.06}_{-0.05}$\\
\vspace{1.5mm}
$q=M_2/M_1$ & -- & $1.245$&$^{+0.038}_{-0.037}$\\
\vspace{1.5mm}
$i$ & deg & $81$&$^{+9}_{-10}$\\
\bottomrule
\end{tabular}
\end{table}

\section{Asteroseismic and spectroscopic analysis}
\label{sec:seismo}

\subsection{Iterative fitting method}

The spectroscopic and asteroseismic analyses are interconnected, because the system metallicity and the luminosity ratio of the stars produce a similar effect on the observed depths of the red giant's spectral lines: both a higher metallicity and a higher red giant luminosity result in deeper spectral lines for the red giant. The luminosity ratio is determined from the asteroseismology, while the red giant's oscillations are sensitive to the metallicity. We therefore used an iterative approach to arrive at the system parameters, which can be summarised as follows:
\begin{enumerate}
\item Determine the red giant's luminosity, age and mass (initially assuming  solar metallicity).
\item Calculate the $\delta$\,Sct mass using the orbital mass ratio.
\item Using evolutionary tracks, find the luminosity of the $\delta$\,Sct star given that age, $\delta$\,Sct mass, and system metallicity.
\item Calculate the light ratio and its uncertainty.
\item Use the new light ratio to refine the spectroscopic metallicity.
\item Repeat from step (i) with the new metallicity.
\end{enumerate}

For the first iteration of step (i), we used the asteroseismic scaling relations with $\Delta\nu=8.05\pm0.04$\,$\upmu$Hz and $\nu_{\rm max}=102.06\pm0.96$\,$\upmu$Hz for the red-giant oscillations. Using figure~1 of \citet{mosseretal2014}, figure~4 of \citet{stelloetal2013a} and figure~2 of \citet{yangetal2012}, we inferred that the red giant is probably a secondary clump star (core He burning) with a mass of $\sim$2.45$\pm$0.11\,M$_{\odot}$ and an age of 500--700\,Myr. Then (ii), using the binary mass ratio $q=1.245\pm0.038$ we inferred a $\delta$\,Sct mass of $1.97\pm0.1$\,M$_{\odot}$. At solar metallicity, this gave a luminosity range of 16.6 to 26.4\,L$_{\odot}$ for the $\delta$\,Sct component (iii), compared with $L_{\rm RG} = 47$\,L$_{\odot}$ from the measured $T_{\rm eff}$ and calculated $R_{\rm RG}$, for a light ratio of $L_{\rm RG}/L_{\rm \delta\,Sct} = 2.3\pm0.5$ (iv).

Using this light ratio, the spectroscopic parameters for the red giant were revised slightly (v), with the metallicity changing by $-0.11$\,dex on the first iteration, and by a total of $+0.02$\,dex on subsequent iterations (vi), to become those given in Table\:\ref{tab:rev-spec} at a light ratio of $L_{\rm RG}/L_{\rm \delta\,Sct} = 3.2$. This value is consistent with the spectroscopic light ratio that results in equal metallicities for the two components. After the first iteration, the asteroseismic modelling used the oscillation frequencies directly rather than relying only on scaling relations, as we now describe.

\begin{table}
\centering
\caption{Spectroscopic parameters for the red giant and $\delta$\,Sct components, after three iterations of asteroseismic inference to determine the luminosity ratio of the stars (see Sect.\,\ref{sec:seismo}). Values are provided for $L_{\rm RG}/L_{\rm \delta\,Sct} = 3.2$.}
\label{tab:rev-spec}
\begin{tabular}{c c r@{}l}
\toprule
\multicolumn{1}{c}{Parameter} & Units & \multicolumn{2}{c}{Values}\\
\midrule
\multicolumn{4}{c}{RG}\\
\midrule
$T_{\rm eff}$ & K & $5124$&$\pm194$\\
$\log g$ & (cgs) & $2.65$&$\pm0.39$\\
$v_{\rm mic}$  & km\,s$^{-1}$ & $0.38$&$\pm0.38$\\
$v_{\rm mac}$  & km\,s$^{-1}$  & $6.1$&$\pm1.2$\\
$v\sin i$ & km\,s$^{-1}$ & $2.5$&$\pm1.5$ \\
${\rm [Fe/H]}$ & & $-0.09$&$\pm0.16$ \\
\midrule
\multicolumn{4}{c}{$\delta$\,Sct}\\
\midrule
$T_{\rm eff}$ & K & $7500$&$\pm210$\\
$\log g$ & (cgs) & $3.7$&$\pm0.6$\\
$v\sin i$ & km\,s$^{-1}$ & $46$&$\pm7$\phantom{0} \\
\bottomrule
\end{tabular}
\end{table}

\subsection{Stellar model calculations}

Asteroseismic modelling of the red giant was carried out with the stellar evolution code {\sc mesa} (Modules for Experiments in Stellar Astrophysics, version 12115, \citealt{paxtonetal2011,paxtonetal2013, paxtonetal2015,paxtonetal2018}) and the oscillation code {\sc gyre} (version 5.1, \citealt{townsend&teitler2013}). We adopted the solar chemical mixture [$(Z/X)_{\odot}$ = 0.0181] provided by \citet{asplundetal2009}. 
The initial chemical composition was calculated by: 
\begin{equation}
\log (Z_{\rm{init}}/X_{\rm{init}}) = \log (Z/X)_{\odot} + \rm{[Fe/H]}.  \\
\end{equation}
We used the {\sc mesa} $\rho-T$ tables based on the 2005 update of the OPAL equation of state tables \citep{rogers&nayfonov2002} and we used OPAL opacities supplemented by the low-temperature opacities from \citet{fergusonetal2005}. The {\sc mesa} `simple' photosphere was used for the set of boundary conditions for modelling the atmosphere; alternative model atmosphere choices do not strongly affect the results for solar-like oscillators \citep{yildiz2007,joyce&chaboyer2018b,nsambaetal2018,vianietal2018} or for $\delta$\,Sct stars \citep{murphyetal2021a}. The mixing-length theory of convection was implemented, where $\alpha_{\rm MLT} = \ell_{\rm MLT}/H_{\rm p}$ is the mixing-length parameter. The exponential scheme by \citet{herwig2000} was adopted for the convective core overshooting, where the diffusion coefficient in the overshoot region is given as  \begin{eqnarray}
{D_{\rm{OV}}} = {D_{\rm{conv,0}}}\exp \left( - \frac{{2(r-r_{0})}}{{(f_0 + f_{\rm{ov}}){H
_{p}}}}\right).
\end{eqnarray}
Here, $D_{\rm{conv,0}}$ is the diffusion coefficient from the mixing-length theory at a user-defined location near the Schwarzschild boundary. The switch from convection to overshooting is set to occur at $r_0$. To consider the step taken inside the convective region, $(f_0 + f_{\rm{ov}}){H_{\rm p}}$ is used. In {\sc mesa}, $f_{\rm{ov}}$ is a free parameter and $f_{\rm 0}$ equals 0.5$f_{\rm{ov}}$. 
We set the overshooting parameter $f_{\rm{ov}}$ = (0.13$M$ - 0.098)/9.0  and adopted a fixed $f_{\rm{ov}}$ at 0.018 for models above $M = 2.0$\,M$_{\odot}$ following the mass-overshooting relation found by \citet{2010ApJ...718.1378M}. We also applied the {\sc mesa} predictive mixing scheme in our model for a smooth convective boundary. The mass-loss rate on the red-giant branch with Reimers prescription is set as $\eta = 0.2$, which is constrained by old open clusters NGC\,6791 and NGC\,6819 \citep{miglioetal2012}.

The model computation ranged from 1.8 to 3.0$M_{\odot}$. 
We computed each stellar evolutionary track from the Hayashi line to the point on the red-giant branch where \mbox{$\log g$ = 1.5\,dex}. Each evolutionary track includes both H-burning (red-giant branch) and He-burning (red-clump) phases. The grid includes four independent model inputs: stellar mass ($M$), initial helium fraction ($Y_{\rm init}$), initial metallicity ([Fe/H]), and the mixing-length parameter ($\alpha_{\rm MLT}$). Ranges and grid steps of the four model inputs are summarized in Table \ref{tab:grid}.

\begin{table}
\centering
\caption{Input ranges and grid steps of the model grid.}
\label{tab:grid}
\begin{tabular}{lccc} % four columns, alignment for each
\toprule
Input Parameter & \multicolumn{2}{c}{Range} & Increment\\
& From & To & \\
\midrule
\multicolumn{4}{c}{Giant component}\\
\midrule
$M / $M$_{\odot}$  & \phantom{$-$}1.80 & 3.00 &  0.02\\
$\rm{[Fe/H]}$ & $-0.4$\phantom{0} & 0.4\phantom{0} & 0.1\phantom{0}\\
$Y_{\rm init}$ & \phantom{$-$}0.24 & 0.32 & 0.02\\
$\alpha_{\rm{MLT}}$ & \phantom{$-$}1.7\phantom{0} & 2.3\phantom{0} & 0.2\phantom{0}\\
\midrule
\multicolumn{4}{c}{$\delta$\,Sct component}\\
\midrule
$M / $M$_{\odot}$  & \phantom{$-$}1.60 & 1.90 &  0.02\\
$\rm{[Fe/H]}$ & $-0.4$\phantom{0} & 0.4\phantom{0} & 0.1\phantom{0}\\
$Y_{\rm init}$ & \phantom{$-$}0.24 & 0.32 & 0.02\\
$\alpha_{\rm{MLT}}$ & \phantom{$-$}1.7\phantom{0} & 2.3\phantom{0} & 0.2\phantom{0}\\
\bottomrule
\end{tabular}
\end{table}

Modelling results in red clump stars can be sensitive to the prescription of convective overshooting \citep[e.g.][]{bossinietal2017}. Our prescription is that demonstrated in figure\:1 of \citep{constantinoetal2015}, who compared model predictions for $\Delta\Pi$ as a function of stellar mass with observations from \citet{mosseretal2014}. 
%For stars having small period spacings for their mass, standard overshooting models such as the one we employ appear to be more appropriate. 
We review the choice of overshooting parameters at the end of this section.

\subsection{Oscillation Models}
\label{ssec:oscillation_models}

Theoretical stellar oscillations for $\ell$ = 0, 1, and 2 were calculated with \textsc{GYRE} (version 5.1; \citealt{townsend&teitler2013}), by solving the adiabatic stellar pulsation equations with the structural models generated by {\sc mesa}. Our computation was carried out in three steps as described below to avoid excessive computing time. 
\begin{itemize}
    \item[1.] Searching for stellar models within a 3$\sigma$ cube constrained by the observed $T_{\rm eff}$, $\log g$, and [Fe/H], calculating a spectroscopic likelihood $L_{\rm Spec}$ ($L = e^{-\chi^{2}}$, where $\chi^{2}$ is the reduced $\chi$ square) for each model.
    \item[2.] Computing radial ($\ell = 0$) mode frequencies for models whose $L_{\rm spec} \geq 0.001$, fitting those frequencies and calculating seismic likelihood for the radial modes ($L_{\rm seis, l =0}$), and then deriving a total likelihood \mbox{$L^{'} = L_{\rm spec} \times L_{\rm seis, l =0}$}. 
    \item[3.] Computing  $\ell = 1$ and 2 mode frequencies for models whose $L^{'}$/$L^{'}_{\rm max} \geq 0.001$, fitting all observed modes and calculating the seismic likelihood for all modes ($L_{\rm seis}$), and lastly deriving a final likelihood with the method described by \citet{tlietal2020} for each model to estimate stellar parameters.
\end{itemize}
For correcting the surface term of the red giant oscillations only, we used the two-term expression described by \citet{ball&gizon2014}. The parameters of the top 10 models for the red giant are given in Table\:\ref{tab:topten}.

%\begin{figure}
%\begin{center}
%\includegraphics[width=0.5\textwidth]{figures/i3_Top_10_models_on_echelle.pdf}\\
%\caption{Top 10 models on the \'Echelle diagram. Filled black symbols are observations, open blue and red symbols represent models at the red-giant branch and red-clump stages.}
%\label{fig:rg_model_echelle}
%\end{center}
%\end{figure}

\begin{table*}
	\centering
	\caption{Stellar parameters of the top 5 models for the RGB and RC scenarios.}
	\label{tab:topten}
	\begin{tabular}{cccccccccc} % four columns, alignment for each
		\toprule
	    Stage & Normalized Likelihood & Mass& Age& $T_{\rm eff}$ & $\log g$ &[Fe/H] &	$R$ & $L$ & $\Delta\Pi$\\
		 & &(M$_{\odot})$& (Gyr)& (K) & (dex) & (dex) &	(R$_{\odot}$) & (L$_{\odot})$ & (s)\\
\hline
RGB &   1.00 &      2.20 &  0.66 &     5095 &  2.930 &    -0.100 &   8.411 &      42.81 &   196.5 \\
RC &   0.98 &      2.053 &  1.10 &     5140 &  2.920 &    -0.084 &   8.221 &      42.39 &   197.4 \\
RGB &   0.60 &      2.06 &  0.71 &     4990 &  2.916 &    -0.100 &   8.273 &      38.13 &   196.6 \\
RGB &   0.57 &      2.24 &  0.59 &     5193 &  2.931 &     0.000 &   8.483 &      47.05 &   196.2 \\
RC &   0.56 &      2.171 &  1.08 &     5033 &  2.926 &     0.117 &   8.395 &      40.64 &   193.1 \\
RC &   0.52 &      1.993 &  1.07 &     5158 &  2.917 &    -0.085 &   8.134 &      42.08 &   193.1 \\
RGB &   0.51 &      2.30 &  0.62 &     5166 &  2.936 &     0.000 &   8.545 &      46.73 &   196.2 \\
RC &   0.49 &      2.272 &  0.85 &     5201 &  2.932 &     0.017 &   8.531 &      47.86 &   193.2 \\
RC &   0.47 &      1.993 &  1.12 &     5086 &  2.916 &    -0.184 &   8.141 &      39.85 &   193.8 \\
RC &   0.46 &      2.373 &  0.71 &     5275 &  2.940 &    -0.181 &   8.642 &      51.96 &   193.5 \\
\bottomrule
	\end{tabular}
\end{table*}

\subsection{Modelling results}
\label{ssec:modelling_results}

We fitted the spectroscopic observations and asteroseismic mode frequencies by using the likelihood functions and the fitting procedure introduced by \citet{tlietal2020}. This fitting method accounts for the systematic offset between observed and model frequencies (which is larger than the observed uncertainties) and also applies a weighting factor depending on the $\nu/\nu_{\rm max}$ ratio for each peak, which together give more sensible probability distributions for stellar parameters. Specifically, we determined the model systematic uncertainty as the median offset between the observations and the best-fitting model in the penultimate iteration, which was 0.04 for $\ell=0$ modes, 0.11 for $\ell=1$, and 0.05 for $\ell=2$. However, unlike in \citet{tlietal2020}, we did not interpolate the model frequencies in this work. In addition, no perturbative or other formulation was applied to the frequencies to account for rotation.

\begin{figure}
\begin{center}
\includegraphics[width=0.50\textwidth]{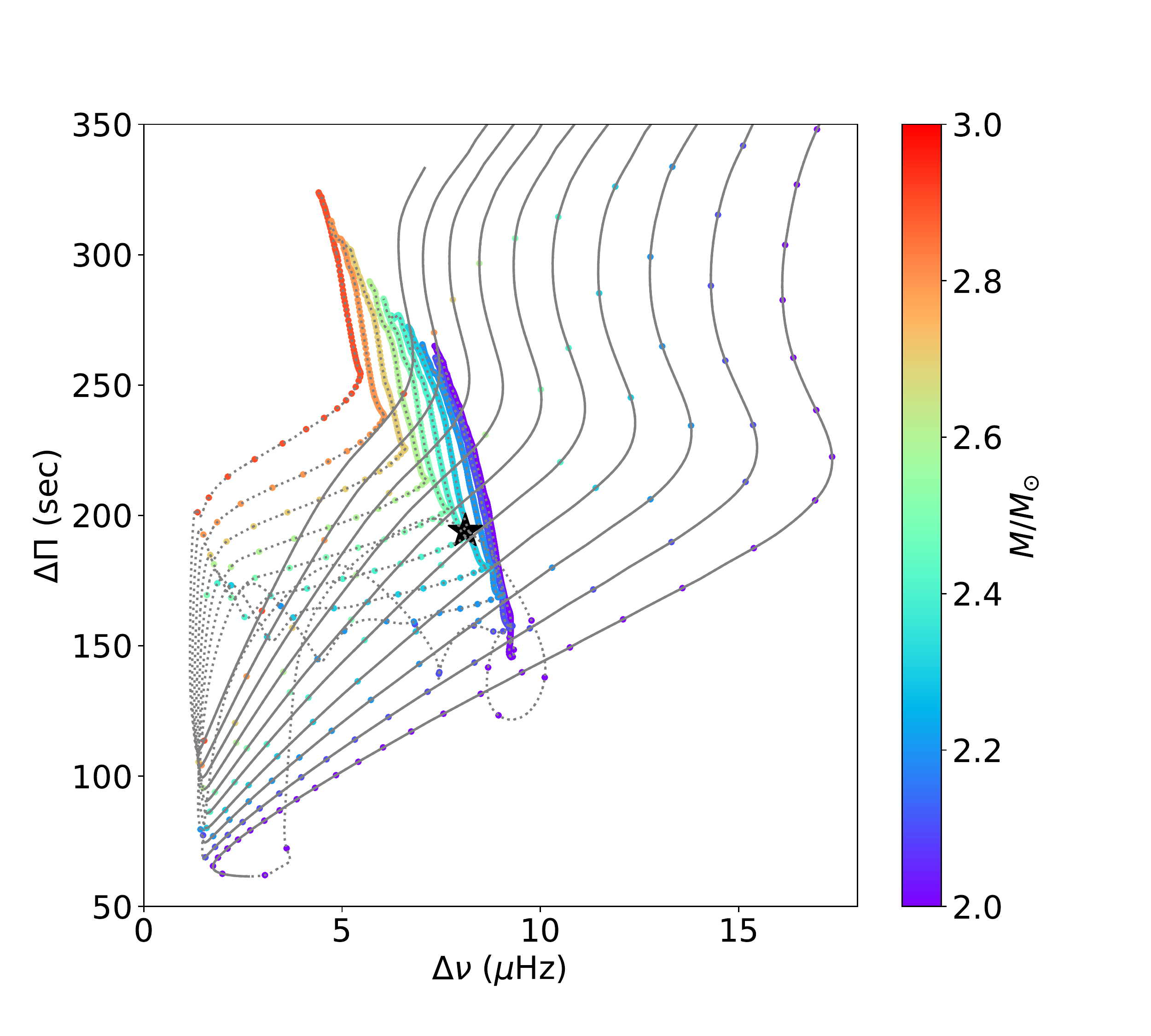}\\
\caption{Determining the evolutionary stage of the red giant component on the $\Delta\nu - \Delta\Pi$ diagram. Solid and dotted lines indicate H-shell-burning (RGB) and He-core-burning (RC) Phases. Evolutionary tracks in this figure range from 2.0 to 2.9\,M$_{\odot}$ in steps of 0.1\,M$_{\odot}$, with [Fe/H] = $-0.1$\,dex. Filled circles are located at regular intervals of 1\,Myr on each track. The black filled 'star' shows the location of the red giant component, compatible with both RGB and RC evolutionary stages.}
\label{fig:evol_phase}
\end{center}
\end{figure}

The modelling results show that the red giant could be an RGB star or a secondary red-clump star (Fig.\,\ref{fig:evol_phase}). The latter is more likely because of its considerably longer duration, with tracks in the middle of the observational error box spending approximately 100 times longer in the secondary RC phase than the RGB phase. While the two phases can be distinguished in lower-mass red giants by the period spacing of their gravity modes, secondary RC stars have similar period spacings to RGB stars. We investigate modelling results for both possibilities.

We start by noting that our measured period spacing (\mga{$\Delta\Pi = 194.0$}\,s; Sec.\,\ref{ssec:rg}) matches very well with the period spacings of our top ten {\sc mesa} models ($\Delta\Pi = 195\pm2$\,s; Table\:\ref{tab:topten}). This indicates that $\Delta\Pi$ values calculated in {\sc mesa} using the integral $\Delta P = \sqrt{2}\uppi^2(\int N/r~dr)^{-1}$, where $N$ is the buoyancy frequency, do not have a significant systematic offset.

\begin{figure}
\begin{center}
\includegraphics[width=0.48\textwidth]{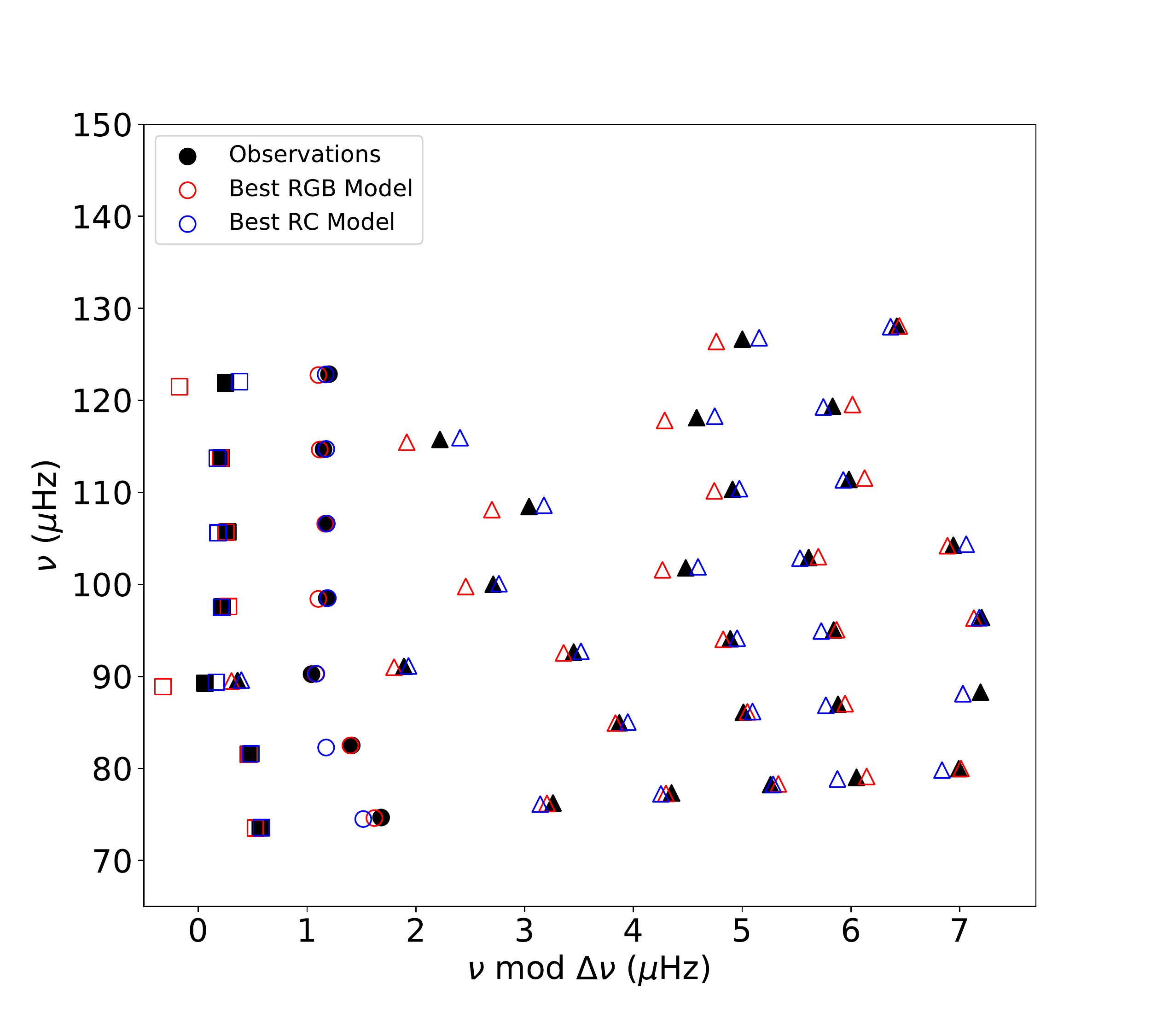}\\
\includegraphics[width=0.42\textwidth]{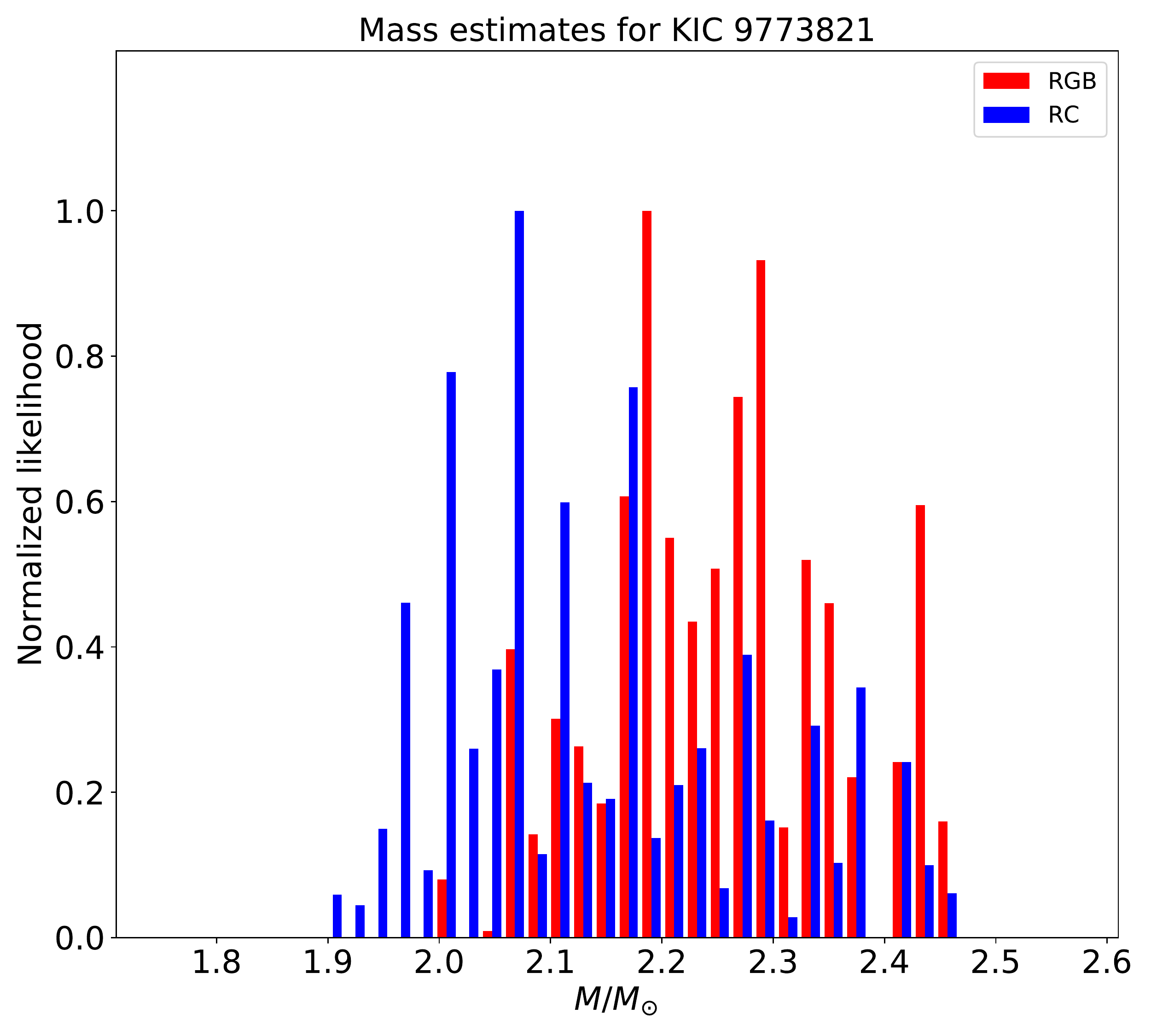}\\
\includegraphics[width=0.42\textwidth]{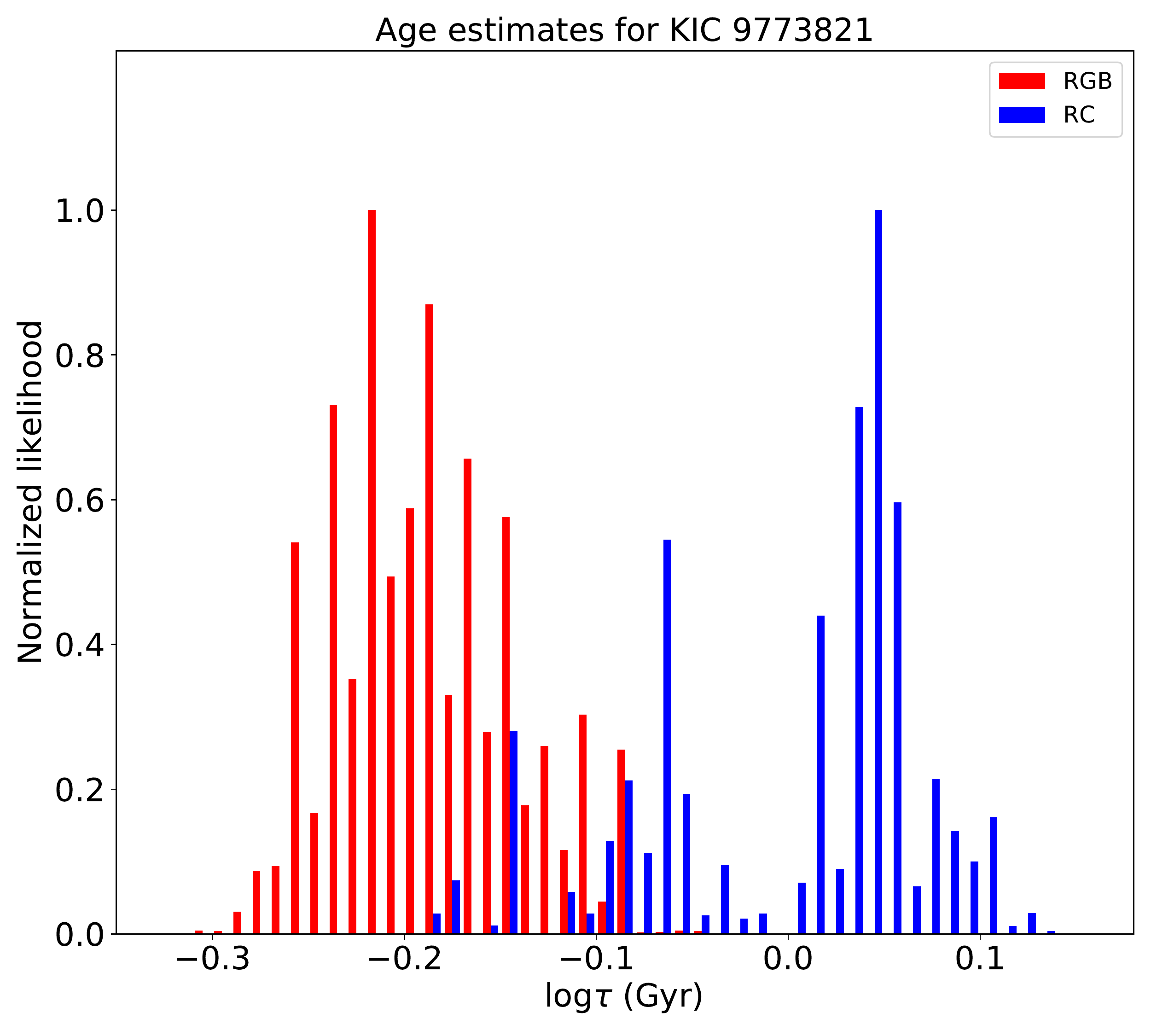}
\caption{{\bf Top:} The observed and modelled frequencies from the best RGB and RC models for KIC\,9773821. Circles show radial ($\ell=0$) modes, triangles show dipole ($\ell=1$) modes, and squares show quadrupole ($\ell=2$) modes. {\bf Middle:} The probability distribution of the modelled red giant mass, normalised to the likelihood of the best value. Although the integrals are similar, the He-burning stage has a much longer duration than H-shell burning, so the former is the more likely evolutionary stage (see the text). {\bf Bottom:} The corresponding distribution in age.}
\label{fig:rg_mass}
\end{center}
\end{figure}

In the final iteration of modelling, the red giant masses were calculated to be 
\mga{$2.26^{+0.10}_{-0.12}$\,M$_{\odot}$ (RGB) and
$2.10^{+0.20}_{-0.10}$\,M$_{\odot}$ (RC), with ages of
$0.64^{+0.08}_{-0.06}$ and 
$1.08^{+0.06}_{-0.24}$\,Gyr,} respectively. The probability distributions of mass and age are shown in Fig.\,\ref{fig:rg_mass}. The metallicity from modelling lies in the range $[{\rm Fe/H}] = -0.1\pm0.2$. 

Under the assumption that this component is a RC star, the binary mass ratio and the RC mass imply a $\delta$\,Sct mass of \mga{$1.69^{+0.11}_{-0.10}$}\,M$_{\odot}$. Using a grid of evolutionary tracks of  $[{\rm Fe/H}] = -0.1$\,dex with mass-intervals of 0.05\,M$_{\odot}$ \citep{murphyetal2019}, we ran a Monte-Carlo simulation to determine the effective temperature and luminosity of the $\delta$\,Sct star and to check the validity of our joint asteroseismic and spectroscopic analysis. 
In this process, we generated 10\,000 random anti-correlated masses and ages from the above distributions, such that larger masses accompanied younger ages, and determined the closest $T_{\rm eff}$ and luminosity from the evolutionary tracks at fixed metallicity. Using the median and standard deviation of the 10\,000 samples, we found that the $\delta$\,Sct star has \mga{$T_{\rm eff}=7820^{+490}_{-120}$}\,K, and
\mga{$L=14.8^{+4.3}_{-3.4}$}\,L$_{\odot}$. 
For a red giant luminosity of 
41$\pm$2\,L$_{\odot}$, this gives a luminosity ratio of
$L_{\rm RG}/L_{\rm \delta\,Sct} = $ \mga{$2.8^{+0.8}_{-0.6}$}, which is consistent with the ratio assumed for the same iteration of spectroscopic analysis ($L_{\rm RG}/L_{\rm \delta\,Sct} = 3.2$). 
%The inferred $T_{\rm eff}$ of the $\delta$\,Sct star is also consistent with the spectroscopic value.
In the final iteration, we found that the red-giant modelling was dominated by the oscillation frequencies: small changes in metallicity of $\sim$0.02\,dex resulting from small changes in the luminosity ratio had little impact on the red giant mass, luminosity or age. The physical parameters of each star from this final iteration are given in Table\:\ref{tab:params}.

We note that if the red giant were on the RGB instead of the secondary clump, then the $\delta$\,Sct star would have a larger mass, at \mga{$M_{\delta\,{\rm Sct}} =1.81\pm0.14$}\,M$_{\odot}$. It would be considerably hotter, with \mga{$T_{\rm eff} = 8700^{+320}_{-440}$}\,K, while the younger age \mga{($0.64^{+0.08}_{-0.06}$\,Gyr)} would make it less evolved and thus only slightly more luminous \mga{($L=16.9^{+6.9}_{-4.9}$\,L$_{\odot}$).}
We note that this inferred temperature is considerably hotter than the spectroscopic constraint (7500\,K), and places the $\delta$\,Sct star close the blue edge of the instability strip. In fact, \mga{59}\% of the sampled positions in our Monte-Carlo process lie outside the theoretical $\delta$\,Sct instability strip in this scenario. We also note that young $\delta$\,Sct stars have much higher frequency oscillations than those observed here \citep{beddingetal2020}, which lends support to the argument that the secondary clump scenario is the correct one.

\subsection{Comparison with the $\nu_{\rm max}$ scaling relation}

Here we investigate whether secondary clump stars follow the widely-used $\nu_{\rm max}$ scaling relation, $\nu_{\rm max} \propto g/\sqrt{T_{\rm eff}}$ \citep{brownetal1991,kjeldsen&bedding1995}, since $\nu_{\rm max}$ was not used in the model calculations. 
We can write this relation as
\begin{equation}
    \nu_{\rm max}/\nu_{\rm max,\odot} \approx \frac{M}{M_{\odot}} \left(\frac{R}{R_{\odot}}\right)^{-2} 
    \left(\frac{T_{\rm eff}}{5777\,{\rm K}}\right)^{-0.5},
\end{equation}
where we adopt $\nu_{\rm max,\odot} = 3090\pm30$\,$\upmu$Hz \citep{huberetal2011}.  In addition to the 1\% uncertainty in $\nu_{\rm max,\odot}$, we also included a systematic uncertainty in the scaling relation of 1.1\%, as recently suggested by \citet{ylietal2020}. 
%This gives a $\nu_{\rm max}$ value for the best-fitting RGB model in Table~\ref{tab:topten} of $101.4 \pm 1.5$\,$\upmu$Hz, which is in good agreement with the measured value of $102.06\pm0.96$\,$\upmu$Hz. 
This gives a $\nu_{\rm max}$ value for the best-fitting RC model in Table~\ref{tab:topten} of \mga{$99.5\pm1.5$}\,$\upmu$Hz, which agrees at 1.7$\sigma$ with the measured value of $102.06\pm0.96$\,$\upmu$Hz. This is a useful confirmation of the $\nu_{\rm max}$ scaling relation for a star in the secondary clump, which is a regime that has not previously been tested in this way.

%\begin{figure}
%\begin{center}
%\includegraphics[width=0.48\textwidth]{figures/dSct_subsolar_both_HRD_v2.pdf}\\
%\caption{Location of the $\delta$\,Sct star on the HR diagram, for the two possible evolutionary states of the red-giant component. Symbol opacity shows the probability of the shown locations. Theoretical and observed $\delta$\,Sct instability strip boundaries come from \citet{dupretetal2004} and \citet{murphyetal2019}, respectively. This indicative analysis was performed for fixed metallicity of [Fe/H] = $-0.1$.}
%\label{fig:dsct_hrd}
%\end{center}
%\end{figure}

\begin{table}
\centering
\caption{Stellar parameters for the $\delta$\,Scuti and red giant components based on iterative spectroscopic and asteroseismic analysis. Parameters are given for the secondary clump scenario, only. The observed asteroseismic quantities were used for the first iteration, after which individual frequency modelling was used. The age and metallicity were determined for the red giant component and applied to the whole system. Quantities from individual red giant models, including $\log g$ and radius, are supplied in Table\:\ref{tab:topten}.}
\label{tab:params}
\begin{tabular}{c c c}
\toprule
\multicolumn{1}{c}{Parameter} & Units & \multicolumn{1}{c}{Values}\\
\toprule
\multicolumn{3}{c}{Observed RG asteroseismic quantities}\\
\midrule
$\Delta\nu$ & $\upmu$Hz & $8.05\pm0.04$ \\
$\nu_{\rm max}$ & $\upmu$Hz & $102.06\pm0.96$ \\
$\Delta\Pi$  & s & $194.0$\\
\midrule
\multicolumn{3}{c}{Iteratively determined stellar quantities}\\
\multicolumn{3}{c}{(secondary clump scenario)}\\
\midrule
$M_{\rm RG}$ & M$_{\sun}$ & $2.10^{+0.20}_{-0.10}$\\
$M_{\delta\,{\rm Sct}}$ & M$_{\sun}$ & $1.69^{+0.11}_{-0.10}$ \\ 
$T_{\rm eff,RG}$ & K & $5124\pm194$ \\
$T_{\rm eff,\delta\,Sct}$ & K & $7820^{+490}_{-120}$\\
$L_{\rm RG}$ & L$_{\sun}$ & $41\pm2$ \\
$L_{\delta\,{\rm Sct}}$ & L$_{\sun}$ & $14.8^{+4.3}_{-3.4}$ \\
\midrule
\multicolumn{3}{c}{Iteratively determined system quantities}\\
\midrule
age & Gyr & $1.08^{+0.06}_{-0.24}$ \\
$[$Fe/H$]$ & (dex) & $-0.09\pm0.16$ \\
\bottomrule
\end{tabular}
\end{table}

\subsection{Convective overshooting schemes}

Here we discuss different convective overshooting schemes for the red giant phase, and justify our choice of `standard overshooting' for modelling the red giant component of KIC\,9773821. According to \citet{constantinoetal2015,constantinoetal2016} and \citet{bossinietal2017}, asymptotic g-mode period spacings predicted by models with the `standard overshooting' scheme (proposed by \citealt{herwig2000}) during the core helium burning (CHeB) lifetime are, on average, smaller than those found in \textit{Kepler} CHeB stars and open clusters. To remedy this, a `maximal-overshoot' scheme has been proposed \citep{constantinoetal2015}, which does not pretend to be realistic but produces the most massive convective core possible. This scheme extends the CHeB lifetime of models and hence increases the period spacing at the end of CHeB phase so that it matches the upper boundary of the observed $\Delta\Pi$ distribution, which is dominated by low-mass stars. On the other hand, the `standard overshooting' scheme is well calibrated with turn-off stars in open clusters and with main-sequence stars in eclipsing binary systems, and better fits more massive stars \citep{deheuvels2020}. However, the calibrations for main-sequence stars could be unsuitable because of the dramatic changes in stellar structure between the main-sequence turn-off and the onset of CHeB.

A relatively comprehensive way to construct evolutionary models for CHeB stars could be to start with the standard overshooting scheme to provide the best accuracy during the main-sequence phase, then switch to the maximal overshooting scheme at some point before the core-helium burning phase. The switch of overshooting scheme necessarily complicates the model computation because it introduces at least two additional free parameters: the optimal time to make the switch, and the `best' scheme to switch to. Given that we find the RG component of KIC\,9773821 to be at the beginning of its CHeB phase, there has been little CHeB in its convective core thus far, and any switch of overshooting scheme would have little time to affect the star's evolution. Retaining standard overshooting is therefore an appropriate approximation in this case. In other words, KIC\,9773821 is not one of the aforementioned low-mass stars at the end of the CHeB phase whose period spacings are better reproduced with maximal overshooting. In figure 1 of \citet{constantinoetal2015}, models with the standard overshooting scheme better predict the lower end of the $\Delta\Pi$ distribution, where high-mass and relatively young CHeB stars such as KIC\,9773821 lie. The adoption of standard overshooting leads to a slightly underestimated age, but only compared to the maximal overshooting model, which is not a more valid choice for this star. Indeed, maximal overshooting is not always a better predictor of the observed period spacings \citep{arentoftetal2017}.

In summary, we estimate that the effect of our choice of standard overshooting is not significant because the time this star spends in the CHeB phase is only a small fraction of the total CHeB duration. The fact that the period spacings we calculate with {\sc mesa} are in good agreement with our observed period spacings supports this.

\section{$\delta$\,Sct mode identification via asteroseismic modelling}
\label{sec:dsct_models}

The orbit and the asteroseismic solution for the red giant component have constrained the mass, age, and location on the HR diagram of the $\delta$\,Sct star. These constraints offer a chance to identify the five $\delta$\,Sct modes (Section~\ref{ssec:dsct}). We modelled the $\delta$\,Sct star with the same theoretical codes and input physics as the red giant. Non-rotating models were computed within a mass range of 1.6--1.9\,$M_{\odot}$. This is despite the fact that $v \sin i$ was measured, because the inclination angle is unknown. While the orbital inclination was measured to be near edge-on and the red-giant inclination is also suggestive of a egde-on inclination, there is no guarantee of spin-orbit alignment. While non-rotating models ignore the possibility that the peaks could be rotationally split non-radial modes, we found satisfactory results considering radial modes and dipole (mixed) modes with $m=0$. Details of the grid computations can be seen in Table~\ref{tab:grid}.

\subsection{The `$\delta$\,Sct+RC' Scenario}

\begin{figure}
\begin{center}
\includegraphics[width=0.47\textwidth]{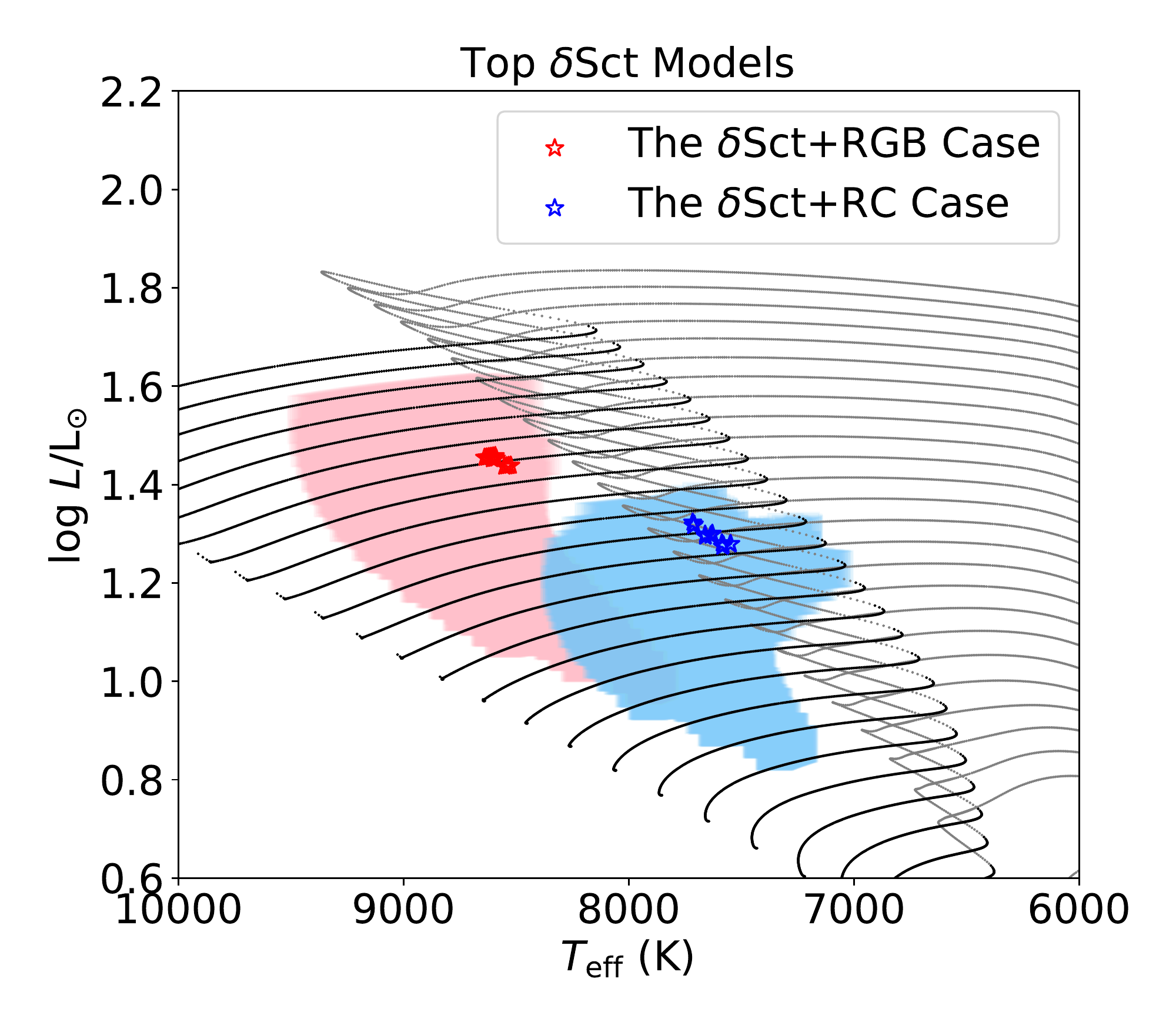}\\
\includegraphics[width=0.47\textwidth]{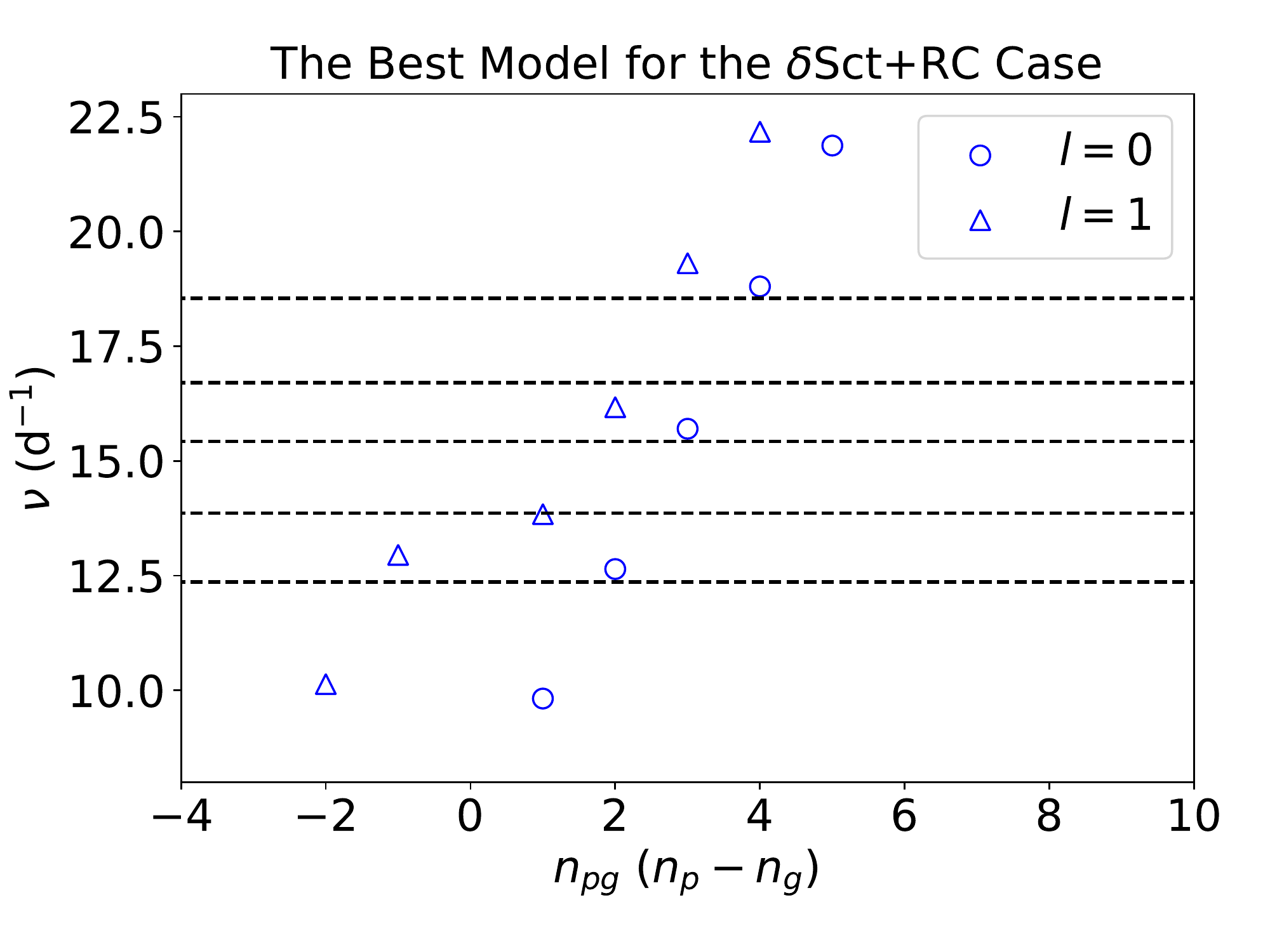}
\includegraphics[width=0.47\textwidth]{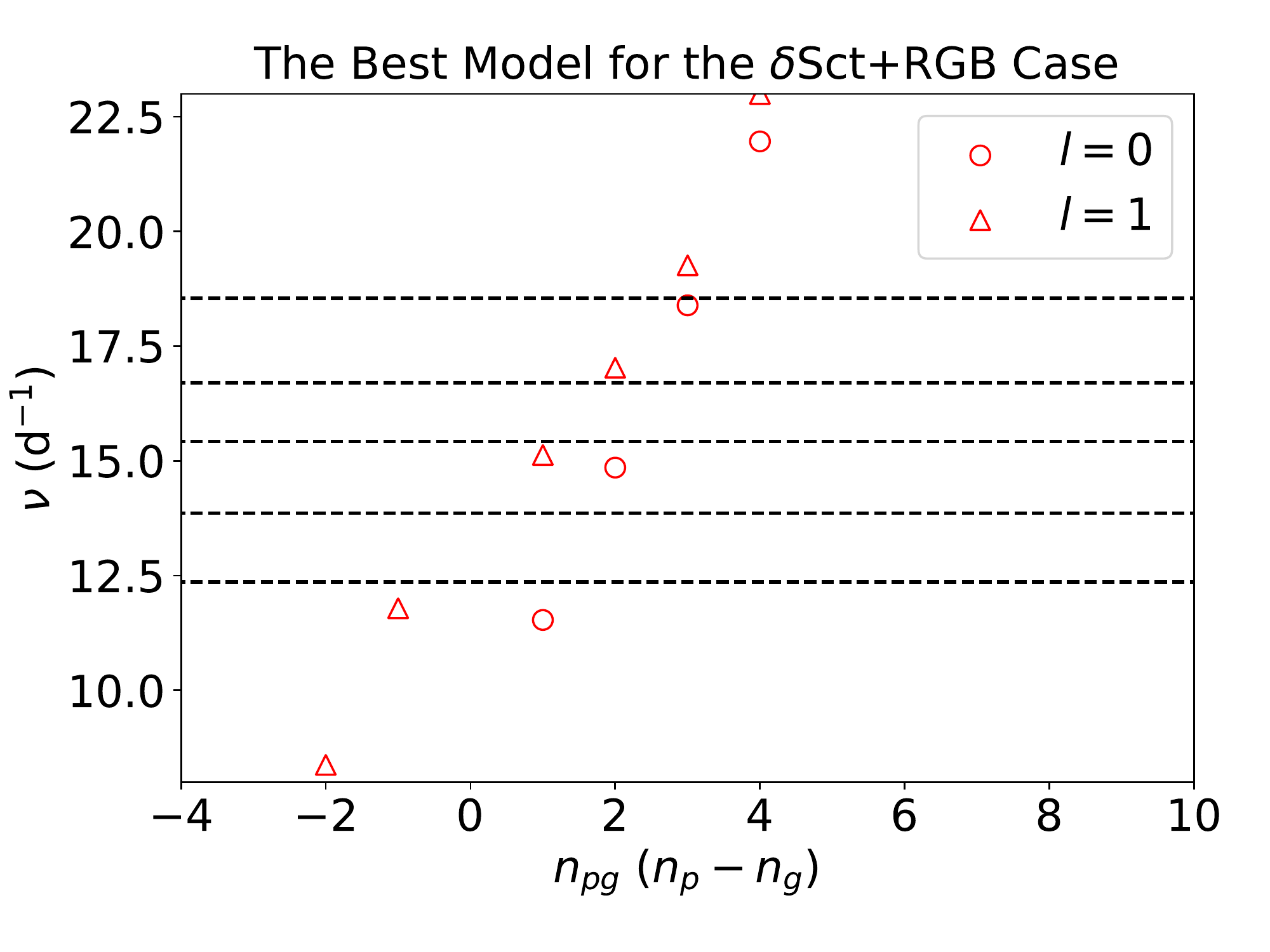}
\caption{{\bf Top:} 
Theoretical models for the $\delta$\,Sct star on the HR diagram. Dots are evolutionary tracks of [Fe/H] $=-0.1$\,dex computed with {\sc mesa}. We use black and grey dots to distinguish models before and after the `hook'. Red and blue stars represent top models for $\delta$\,Sct+RGB and $\delta$\,Sct+RC scenarios, respectively. The red and light blue shading indicate the region of all $\delta$\,Sct models examined. The top ten $\delta$\,Sct models in each case happen to lie before the `hook'. {\bf Middle:} 
The best $\delta$\,Sct model for the $\delta$\,Sct+RC scenario. 
Black dashed lines are the five $\delta$\,Sct peaks listed in Table~\ref{tab:freqs}. Open circles and triangles represent theoretical frequencies for $\ell$ = 0 and 1. {\bf Bottom:} As the middle panel, but for the $\delta$\,Sct+RGB scenario.}
\label{fig:dsct_models}
\end{center}
\end{figure}

We first studied the `$\delta$\,Sct+RC' scenario, where the red giant is core-He burning. According to the constraints from the RC models (Sec.\,\ref{ssec:modelling_results}), we set hard limits for mass and age for the $\delta$\,Sct component as \mga{$1.69\pm0.11$\,M$_{\odot}$ and $1.08^{+0.06}_{-0.24}$\,Gyr.} All $\delta$\,Sct models are shown in the top panel of Fig.\,\ref{fig:dsct_models} by blue shading, and all are located on the main sequence before the overall contraction phase (`hook'). Details on the use of {\sc gyre} to compute radial and dipole mode frequencies were given in Sec.\,\ref{ssec:oscillation_models}. We fitted each peak to the nearest theoretical frequency of each model without any prior, allowing each peak to have either $\ell$ = 0 or $\ell$ = 1. We then used a maximum likelihood estimation method to find the best-fitting models by minimising the differences between the observed and modelled frequencies. All of the top 10 models have masses from 1.62 to 1.66\,M$_{\odot}$ and [Fe/H]$ = -0.2$, which are consistent with the observed constraints at the $1\sigma$ level. The best-fitting model in this scenario has the following properties: \mga{$M = 1.62$\,M$_{\odot}$, 
$\tau$ = 1.10\,Gyr, 
$T_{\rm eff}$ = 7581\,K, $\log g$ = 3.84, $L$ = 18.9\,${\rm L_{\odot}}$, and [Fe/H] = $-0.2$}. In the middle panel of Fig.\,\ref{fig:dsct_models}, we show a comparison of the oscillation frequencies of this model with the observations. The observed peaks are well fitted except for $f_4$ at 16.7\,d$^{-1}$. The mode identifications of these 10 models are consistent: $f_1$ (at 12.35\,d$^{-1}$) is the first radial overtone ($\ell$ = 0, $n_p$ = 2); $f_2$ (15.43\,d$^{-1}$) is the second radial overtone ($\ell$ = 0, $n_p$ = 3), $f_5$ (18.54\,d$^{-1}$) is the third radial overtone ($\ell$ = 0, $n_p$ = 4), and $f_3$ (13.86\,d$^{-1}$) is a dipole mixed mode ($\ell$ = 1 $n_{pg}$ = 1). The fundamental radial mode ($\ell=0$, $n=1$) is predicted to have a frequency of 9.734\,d$^{-1}$, which we do not observe. This falls in the region of the red giant oscillations and must have a very low amplitude if it is excited at all.

The model frequencies may have small offsets from a hypothetical rotating model because our models are non-rotating (see, e.g. \citealt{dicriscienzoetal2008}), which may explain some of the small differences between the calculated and the observed frequencies.

\subsection{The `$\delta$\,Sct+RGB' Scenario}

We also studied the `$\delta$\,Sct+RGB' scenario with the same method. The parameter limits for mass and age were \mga{$1.81\pm0.14$, and $0.64^{+0.08}_{-0.06}$\,Gyr.} All examined $\delta$\,Sct models are shown with red shading in the top panel of Fig.\,\ref{fig:dsct_models}. Compared with the $\delta$\,Sct+RC scenario, these models are at earlier evolutionary stages because the $\delta$\,Sct+RGB system age is younger. The best-fitting model in this scenario has \mga{$M$ = 1.92\,M$_{\odot}$, 
$\tau$ = 0.67\,Gyr, 
$T_{\rm eff}$ = 8622\,K, $\log g$ = 3.961, $L$ = 28.6\,${\rm L_{\odot}}$, and [Fe/H] = 0.0}, and is illustrated in the bottom panel of Fig.\,\ref{fig:dsct_models}.
Even in the best model there is not good agreement with observations, with only one mode being unambiguously matched to a reasonable tolerance ($f_5=18.54$\,d$^{-1}$ as the $n=3$ radial mode). The $\chi^2$ of the best-fitting frequencies for $f_1$ to $f_5$ is three times higher in the $\delta$\,Sct+RGB scenario than the $\delta$\,Sct+RC scenario. Combined with the time-frame and $T_{\rm eff}$ arguments we made previously, we therefore conclude that the $\delta$\,Sct+RGB scenario does not explain the observations well.

\section{Conclusions}
\label{sec:conclusions}

We have used spectroscopic radial velocities and pulsation timing to determine that the two pulsators visible in the {\it Kepler} light curve of KIC\,9773821 are in fact a bound system consisting of a secondary clump star and a main-sequence $\delta$\,Sct star -- the first double-pulsator binary of its kind. We have determined the orbital parameters, which include a period of 481.9\,d, an eccentricity of 0.241, and a mass ratio $q = M_{\delta {\rm Sct}}/M_{\rm RG} = 0.80$.

An iterative procedure involving frequency modelling and spectroscopic atmospheric parameter determination has constrained the evolutionary properties of the system. We found the system to be slightly metal poor, with [Fe/H] $\sim$ $-0.1$, and we concluded that the red giant is a secondary clump (core-He burning) star with a mass of \mga{$2.10^{+0.20}_{-0.10}$\,M$_{\sun}$} and an age of \mga{$1.08^{+0.06}_{-0.24}$\,Gyr} (1$\sigma$ uncertainties). These constraints have also allowed four of the five modes of the $\delta$\,Sct star to be identified as a mixed $\ell=1$ mode and three radial modes. Thus the combination of two different types of pulsator in one system has facilitated deeper analyses than would normally be possible in isolation.

We give three arguments to suggest that the red giant is a red clump (RC) star, rather than a star ascending the red giant branch (RGB): (i) the RC timescale is two orders of magnitude longer, so this phase is more likely to be observed; (ii) the corresponding properties of the $\delta$\,Sct star are more consistent with this; and (iii) pulsation models for the younger (RGB) scenario are a poor fit to the $\delta$\,Sct oscillations, whereas the older (RC) scenario fits very well with only simple assumptions.

The orbital inclination was found to be edge on, and a projected rotational velocity $v \sin i = 46\pm7$\,km\,s$^{-1}$ for the $\delta$\,Sct star was measured, but our pulsation models did not include rotation because the potential for spin-orbit misalignment precludes independent measurement of the equatorial rotation velocity. The observed frequencies are well described with radial modes nonetheless, and future exploration of rotating models of a range of inclinations remains possible.

Our precision modelling of a low-mass secondary clump star (i.e. having a mass not much greater than 1.8\,M$_{\sun}$) has offered observational tests in a regime that has not been well studied previously. For instance, we have been able to confirm the $\nu_{\rm max}$ scaling relation for a secondary clump star for the first time, and that period spacings calculated in {\sc mesa} are accurate.

\section*{Acknowledgements}

S.J.M. was supported by the Australian Research Council through DECRA DE180101104 and DP210103119. T.L. acknowledges the funding from the European Research Council (ERC) under the European Union Horizon 2020 research and innovation programme (CartographY GA. 804752). D.H. acknowledges support from the Alfred P. Sloan Foundation and the National Aeronautics and Space Administration (80NSSC19K0597), and the National Science Foundation (AST-1717000). The research leading to these results has received funding from the Fonds Wetenschappelijk Onderzoek - Vlaanderen (FWO) under grant G0H5416N (ERC Opvangproject).

\section{Data Availability}
The pulsation frequencies and mode IDs for the red giant are provided in Table\:\ref{tab:RG_freqs}. Radial velocity data are given in Table\:\ref{tab:RG_RVs}. Time delays for the $\delta$\,Sct component are made available as supplementary online information.

%%%%%%%%%%%%%%%%%%%%%%%%%%%%%%%%%%%%%%%%%%%%%

%%%%%%%%%%%%%%%%%%%% REFERENCES %%%%%%%%%%%%%%%%%

\bibliographystyle{mnras}
\interlinepenalty=10000
\bibliography{sjm_bibliography,tanda,tim,Sanjay} % name of .bib file, without extension

\appendix

\section{Data tables}

The pulsation frequencies and mode identifications for the red giant are given in Table\:\ref{tab:RG_freqs}. Radial velocities for the red giant component are given in Table\:\ref{tab:RG_RVs}. A sample of the $\delta$\,Sct time delays (supplementary online information) is provided in Table\:\ref{tab:TDs}.

\begin{table}
\centering
\caption{Pulsation frequencies and mode degrees from the red-giant component. } %radial orders have been confirmed
\label{tab:RG_freqs}
\begin{tabular}{c c c r}
\toprule
\multicolumn{1}{c}{Frequency} & Uncertainty & Degree & \multicolumn{1}{c}{Order}\\
\multicolumn{1}{c}{$\upmu$Hz} & $\upmu$Hz & ($\ell$) & ($n$)\\
\midrule
$\phantom{1}	73.56	$&$	0.05	$&$	2	$&$		$\\
$\phantom{1}	74.67	$&$	0.05	$&$	0	$&$	8	$\\
$\phantom{1}	76.25	$&$	0.05	$&$	1	$&$		$\\
$\phantom{1}	77.34	$&$	0.05	$&$	1	$&$		$\\
$\phantom{1}	78.25	$&$	0.05	$&$	1	$&$		$\\
$\phantom{1}	79.04	$&$	0.05	$&$	1	$&$		$\\
$\phantom{1}	79.98	$&$	0.05	$&$	1	$&$		$\\
$\phantom{1}	81.56	$&$	0.05	$&$	2	$&$		$\\
$\phantom{1}	82.51	$&$	0.05	$&$	0	$&$	9	$\\
$\phantom{1}	84.97	$&$	0.05	$&$	1	$&$		$\\
$\phantom{1}	86.11	$&$	0.05	$&$	1	$&$		$\\
$\phantom{1}	86.98	$&$	0.05	$&$	1	$&$		$\\
$\phantom{1}	88.29	$&$	0.05	$&$	1	$&$		$\\
$\phantom{1}	89.27	$&$	0.05	$&$	2	$&$		$\\
$\phantom{1}	89.57	$&$	0.05	$&$	1	$&$		$\\
$\phantom{1}	90.25	$&$	0.05	$&$	0	$&$	10	$\\
$\phantom{1}	91.10	$&$	0.05	$&$	1	$&$		$\\
$\phantom{1}	92.66	$&$	0.05	$&$	1	$&$		$\\
$\phantom{1}	94.10	$&$	0.05	$&$	1	$&$		$\\
$\phantom{1}	95.05	$&$	0.05	$&$	1	$&$		$\\
$\phantom{1}	96.41	$&$	0.05	$&$	1	$&$		$\\
$\phantom{1}	97.54	$&$	0.05	$&$	2	$&$		$\\
$\phantom{1}	98.50	$&$	0.05	$&$	0	$&$	11	$\\
$	100.03	$&$	0.05	$&$	1	$&$		$\\
$	101.80	$&$	0.05	$&$	1	$&$		$\\
$	102.93	$&$	0.03	$&$	1	$&$		$\\
$	104.26	$&$	0.05	$&$	1	$&$		$\\
$	105.70	$&$	0.05	$&$	2	$&$		$\\
$	106.61	$&$	0.05	$&$	0	$&$	12	$\\
$	108.47	$&$	0.05	$&$	1	$&$		$\\
$	110.34	$&$	0.05	$&$	1	$&$		$\\
$	111.41	$&$	0.05	$&$	1	$&$		$\\
$	113.75	$&$	0.05	$&$	2	$&$		$\\
$	114.69	$&$	0.03	$&$	0	$&$	13	$\\
$	115.76	$&$	0.05	$&$	1	$&$		$\\
$	118.12	$&$	0.05	$&$	1	$&$		$\\
$	119.37	$&$	0.05	$&$	1	$&$		$\\
$	121.90	$&$	0.10	$&$	2	$&$		$\\
$	122.85	$&$	0.05	$&$	0	$&$	14	$\\
$	126.65	$&$	0.05	$&$	1	$&$		$\\
$	128.07	$&$	0.05	$&$	1	$&$		$\\
\bottomrule
\end{tabular}
\end{table}

\begin{table}
\centering
\caption{Radial velocities for the red-giant component of the system.}
\label{tab:RG_RVs}
\begin{tabular}{r r r}
\toprule
\multicolumn{1}{c}{Time (BJD)} & \multicolumn{1}{c}{RV} &  \multicolumn{1}{c}{$\sigma$RV} \\
\multicolumn{1}{c}{d} & km\,s$^{-1}$ & km\,s$^{-1}$ \\
\midrule
$ 2456122.702771 $&$ 25.180 $&$ 0.150 $\\
$ 2456123.612354 $&$ 25.125 $&$ 0.155 $\\
$ 2456173.369422 $&$ 18.490 $&$ 0.160 $\\
$ 2456173.390837 $&$ 18.450 $&$ 0.160 $\\
$ 2458392.378968 $&$ 14.915 $&$ 0.155 $\\
$ 2458405.440501 $&$ 16.650 $&$ 0.160 $\\
$ 2458425.324385 $&$ 19.150 $&$ 0.160 $\\
$ 2458592.710566 $&$ 15.840 $&$ 0.150 $\\
$ 2458601.707377 $&$ 13.160 $&$ 0.160 $\\
$ 2458629.610681 $&$ 2.500 $&$ 0.150  $\\
$ 2458653.547702 $&$ -6.400 $&$ 0.150 $\\
$ 2458680.465544 $&$ -12.200 $&$ 0.160$\\
$ 2458688.562511 $&$ -12.755 $&$ 0.165$\\
$ 2458701.474174 $&$ -12.840 $&$ 0.160$\\
$ 2458710.457154 $&$ -12.330 $&$ 0.150$\\
$ 2458724.422518 $&$ -10.785 $&$ 0.155$\\
$ 2458753.533187 $&$ -5.995 $&$ 0.155 $\\
\bottomrule
\end{tabular}
\end{table}

\begin{table}
\centering
\caption{Weighted-average light arrival-time delays (TDs) for the $\delta$\,Sct component of the system. Only the first ten rows are shown. The full table is available online.}
\label{tab:TDs}
\begin{tabular}{r r r}
\toprule
\multicolumn{1}{c}{Time (BJD)} & \multicolumn{1}{c}{TD} &  \multicolumn{1}{c}{$\sigma$TD} \\
\multicolumn{1}{c}{d} & \multicolumn{1}{c}{s} & \multicolumn{1}{c}{s} \\
\midrule
$2454958.39198$ & $350.7$ & $142.6$\\
$2454969.50828$ & $305.2$ & $143.5$\\
$2454979.52111$ & $444.2$ & $140.7$\\
$2454989.53392$ & $355.1$ & $153.5$\\
$2454999.54669$ & $434.7$ & $168.4$\\
$2455009.53899$ & $385.1$ & $127.4$\\
$2455021.71768$ & $282.8$ & $140.4$\\
$2455031.73030$ & $324.5$ & $130.8$\\
$2455041.74286$ & $204.2$ & $127.2$\\
$2455051.61232$ & $174.2$ & $115.6$\\
\bottomrule
\end{tabular}
\end{table}

%%%%%%%%%%%%%%%%%%%%%%%%%%%%%%%%%%%%%%%%%%%%%%

% endmatter

\bsp	% typesetting comment
\label{lastpage}
\end{document}